\documentclass[]{pasj02}
\usepackage[switch,mathlines]{lineno} 
\usepackage{xcolor}
\jyear{2024}
\Received{}
\Accepted{}
 
\def\be{\begin{equation}}\def\ee{\end{equation}}\def\vlsr{v_{\rm LSR}} \def\Vlsr{\vlsr} 
\def\deg{^\circ}  
\def\vrot{V_{\rm rot}}              \def\kms{km s$^{-1}$}       
      \def\ekms{~{\rm \ km \ s^{-1}}~}  \def\epc{{\rm \ pc} }   \def\apj{ApJ} \def\aap{A\&A} \def\mnras{MNRAS} \def\pasj{PASJ} \def\aj{AJ}   \def\log{{\rm log}}       

\def\kms{km s$^{-1}$}  \def\deg{^\circ}         \def\log{{\rm log}}
    \def\H{{\rm H}} \def\bc{\begin{center}}\def\ec{\end{center}}

 \def\x{\times}          
\def\Te{T_{\rm e}}\def\({\left(} \def\){\right)}\def\[{\left[} \def\]{\right]}        
\def\ss{\subsection}\def\sss{\subsubsection}

\def\cs{CS ($J=2-1$)}

\def\sgrastar{Sgr A$^*$} 

\def\Jybeam{Jy beam$^{-1}$}

\def\G02{G+0.02-0.02+100}

\def\ekpc{{\rm kpc}}

 \def\lw{\linewidth}
 \def\be{\begin{equation}} \def\ee{\end{equation}}
 \def\nH2{n_{\rm H_2}}
 \def\epc{{\rm pc}} 
  
 \def\h40{H40$\alpha$} 
\def\lw{\linewidth}
\def\Te{T_{\rm e}}
\def\emunit{pc cm$^{-6}$}
\def\TC{T_{\rm C}} \def\IC{I_{\rm C}} \def\TL{T_{\rm L}}\def\IL{I_{\rm L}}
\def\e6{\times 10^6} \def\e7{\times 10^7}
\def\emu{pc cm$^{-6}$ }
\def\Te{T_{\rm e}} \def\emunit{pc cm$^{-6}$ }
\def\TK{\left(\Te\over{\rm K}\right)} \def\nuG{\left(\nu\over{\rm GHz}\right)} \def\eJybeam{{\rm Jy~beam^{-1}}}
\def\Te{T_{\rm e}} \def\TC{T_{\rm C}} \def\TL{T_{\rm L}} \def\Ne{n_{\rm e}} \def\IC{I_{\rm C}}
\def\A{\mathcal{A}}
\def\deg{^{\circ}}
\def\star{\sgrastar}

\def\MC{M$_{\rm C}$}

\def\Jybeam{Jy beam$^{-1}$}
\def\mum{$\mu$m}
\def\Tav{\left<\Te\right>} 
\def\Tcmz{\left<T_{\rm e}\right>_{\rm CMZ}} 
\def\Tcmztext{$\Tcmz= 5872 \pm 78 ~{\rm (SE)} ~\pm 3682~{\rm (SD)}$ K } 

\begin{document} 
\title{Electron temperature and emission measure of HII regions in the central molecular zone (CMZ) from \h40 recombination line and continuum emissions by ALMA CMZ Exploration Survey - ACES -  
}
		
	\def\afmark{\altaffilmark}		
	\def\aftext{\altaffiltext} 	
	\def\scr{\scriptsize}  	
	\author{		
	Yoshiaki \textsc{Sofue},\afmark{1}\orcid{0000-0002-4268-6499}\email{sofue@ioa.s.u-tokyo.ac.jp}	\aftext{1}{Institute of Astronomy, The University of Tokyo, Mitaka, Tokyo 181-0015, Japan} 
	\orcid{0000-0001-6353-0170}{Steven~N. \textsc{Longmore}},\afmark{2,3} \aftext{2}{Astrophysics Research Institute, Liverpool John Moores University, IC2, Liverpool Science Park, 146 Brownlow Hill, Liverpool L3 5RF, UK}	
	\aftext{3}{CCosmic Origins Of Life (COOL) , https://coolresearch.io} 
	\orcid{0000-0001-7330-8856}{Daniel \textsc{Walker}},\afmark{4} \aftext{4}{UK ALMA Regional Centre Node, Jodrell Bank Centre for Astrophysics, Oxford Road, The University of Manchester, Manchester M13 9PL, United Kingdom} 
	\orcid{0000-0001-6431-9633}{Adam \textsc{Ginsburg}},\afmark{5} \aftext{5}{Department of Astronomy, University of Florida, P.O. Box 112055, Gainesville, FL 32611}	
	\orcid{0000-0001-9656-7682}{Jonathan D. \textsc{Henshaw}},\afmark{2,6} \aftext{6}{Max Planck Institute for Astronomy, K\"{o}nigstuhl 17, D-69117 Heidelberg, Germany}	
	\orcid{0000-0001-8135-6612}{John \textsc{Bally}},\afmark{7} \aftext{7}{Center for Astrophysics and Space Astronomy; Department of Astrophysical and Planetary Sciences; University of Colorado, Boulder, CO 80389, USA}	
	\orcid{0000-0003-0410-4504}{Ashley T. \textsc{Barnes}},\afmark{8} \aftext{8}{European Southern Observatory (ESO), Karl-Schwarzschild-Strasse 2, D-85748 Garching, Germany}	
	\orcid{0000-0002-6073-9320}{Cara \textsc{Battersby}},\afmark{9} \aftext{9}{Department of Physics, University of Connecticut, 196A Auditorium Road, Unit 3046, Storrs, CT 06269, USA}	
	\orcid{0000-0001-8064-6394}{Laura \textsc{Colzi}},\afmark{10} \aftext{10}{Centro de Astrobiolog{\'\i}a (CAB), CSIC-INTA, Carretera de Ajalvir km 4, Torrej{\'o}n de Ardoz, 28850 Madrid, Spain}	
	Paul \textsc{Ho},\afmark{11} \aftext{11}{Inst. of Astron. and Astrophys., Academia Sinica, AS/NTU Astronomy-Mathematics Building,  Roosevelt Rd, Taipei 10617, Taiwan}	\orcid{0000-0002-3412-4306}		
	Izaskun \textsc{Jimenez}-\textsc{Serra},\afmark{10} 
	\orcid{0000-0003-4493-8714}		
	Elizabeth \textsc{Mills},\afmark{12} \aftext{12}{Department of Physics and Astronomy, University of Kansas, 1251 Wescoe Hall Drive, Lawrence, KS 66045, USA}	\orcid{0000-0001-8782-1992}		
	\orcid{0000-0002-6362-8159}Maya~A. \textsc{Petkova},\afmark{13} \aftext{13}{Space, Earth and Environment Department, Chalmers University of Technology, SE-412 96 Gothenburg, Sweden}	
	\orcid{0000-0001-6113-6241}{Mattia~C. \textsc{Sormani}},\afmark{14} \aftext{14}{Como Lake centre for AstroPhysics (CLAP), DiSAT, Universit{\`a} dell'Insubria, via Valleggio 11, 22100 Como, Italy}	
	\orcid{https://orcid.org/0009-0002-7459-4174}Jennifer \textsc{Wallace},\afmark{9} 
	\orcid{0000-0002-9483-7164}Robin~G. \textsc{Tress},\afmark{15} \aftext{15}{ Institute of Physics, Laboratory for Galaxy Evolution and Spectral Modelling, EPFL, Observatoire de Sauverny, Chemin Pegasi 51, 1290 Versoix, Switzerland}	
	%
	\orcid{0000-0002-0533-8575}{Nazar \textsc{Budaiev}},\afmark{5} 
	\orcid{0009-0004-0685-7678}{Rojita \textsc{Buddhacharya}},\afmark{2,,16} \aftext{16}{Center for Astrophysics $\vert$ Harvard \& Smithsonian, 60 Garden Street, Cambridge, MA, 02138, USA}	
	%
	\orcid{0009-0004-0121-1560}{Christoph Federrath},\afmark{17} \aftext{17}{Research School of Astronomy and Astrophysics, Australian National University, Canberra, ACT2611, Australia}	
	\orcid{0009-0004-0121-1560}{Zi-Xuan \textsc{Feng}},\afmark{14,18}
	\aftext{18}{Universit\"{a}t Heidelberg, Zentrum f\"{u}r Astronomie, Institut f\"{u}r Theoretische Astrophysik, Albert-Ueberle-Str.\ 2, 69120 Heidelberg, Germany}	
	%
	\orcid{0000-0002-8586-6721}{Pablo \textsc{Garc\'ia}},\afmark{19,20} \aftext{19}{Chinese Academy of Sciences South America Center for Astronomy, National Astronomical Observatories, CAS, Beijing 100101, China}	
	\aftext{20}{Instituto de Astronom\'ia, Universidad Cat\'olica del Norte, Av. Angamos 0610, Antofagasta, Chile}	
	\orcid{0000-0002-1313-429X}{Savannah \textsc{Gramze}},\afmark{5} 
	\orcid{0000-0002-7495-4005}{Christian \textsc{Henkel}},\afmark{21} \aftext{21}{MPIfR, Auf dem H\"ugel 69, Bonn, Germany}	
	\orcid{0000-0001-9155-3978}{Pei-Ying \textsc{Hsieh}},\afmark{22} \aftext{22}{National Astronomical Observatory of Japan, 2-21-1 Osawa, Mitaka, Tokyo 181-8588, Japan}	
	\orcid{0000-0001-5950-1932}{Fengwei \textsc{Xu}},\afmark{6} 
	\orcid{0000-0003-4140-5138}{Katharina \textsc{Immer}},\afmark{8} 
	\orcid{0000-0002-5776-9473}{Dani \textsc{Lipman}},\afmark{9} 
	\orcid{0000-0002-0500-4700}{Dariusz C. Lis},\afmark{23} \aftext{23}{Jet Propulsion Laboratory, California Institute of Technology, 4800 Oak Grove Drive, Pasadena, CA 91109, USA}	
	\orcid{0000-0002-3078-9482}{\'Alvaro S\'anchez-Monge},\afmark{24,25} \aftext{24}{Institut de Ci\`encies de l'Espai (ICE), CSIC, Campus UAB, Carrer de Can Magrans s/n, E-08193, Bellaterra, Barcelona, Spain}	
	\aftext{25}{Institut d'Estudis Espacials de Catalunya (IEEC), E-08860, Castelldefels, Barcelona, Spain}	
	Xing \textsc{Lu},\afmark{26} \aftext{26}{Shanghai Astronomical Observatory, Chinese Academy of Sciences, 80 Nandan Road, Shanghai 200030, P.\ R.\ China} \orcid{0000-0003-2619-9305}	
	Mark~R. \textsc{Morris},\afmark{27} \aftext{27}{Department of Physics and Astronomy, University of California, Los Angeles, CA 90095, USA} \orcid{0000-0002-6753-2066} 	
	Francisco \textsc{Nogueras-Lara},\afmark{28} \aftext{28}{Instituto de Astrof\'isica de Andaluc\'ia (CSIC), Glorieta de la Astronom\'ia s/n, E-18008 Granada, Spain} \orcid{0000-0002-6379-7593}	
	\orcid{0000-0001-8224-1956}{J\"urgen \textsc{Ott}},\afmark{29} \aftext{29}{National Radio Astronomy Observatory, P.O. Box O, 1011 Lopezville Road, Socorro, NM 87801, USA} 	
	\orcid{0000-0002-5811-0136}{Dylan M. Par\'e},\afmark{30,31} \aftext{30}{Joint ALMA Observatory, Alonso de Cordova 3107, Vitacura, Casilla 19001, Santiago de Chile, Chile}	
	\aftext{31}{National Radio Astronomy Observatory, 520 Edgemont Road, Charlottesville, VA 22903, USA}	
	\orcid{0000-0002-3972-1978}{Jaime~E. \textsc{Pineda}},\afmark{32} \aftext{32}{MPI for Extraterrestrial Physics, Giessenbachstr. 1, D-85748, Garching, Germany}	
	\orcid{0000-0002-2887-5859}{V{\'\i}ctor M. \textsc{Rivilla}},\afmark{10} 
	\orcid{0000-0003-3341-6144}{Jairo \textsc{Armijos-Abenda\~no}},\afmark{33} \aftext{33}{Observatorio Astron\'omico de Quito, Observatorio Nacional, Escuela Polit\'ecnica Nacional, Interior del Parque La Alameda, 170403, Quito, Ecuador}	
	\orcid{0000-0003-2384-6589}{Qizhou Zhang},\afmark{16}
	\orcid{0000-0001-5389-0535}{Denise \textsc{Riquelme-V\'asquez}},\afmark{34} \aftext{34}{Departamento de Astronom\'ia, Universidad de La Serena, Ra\'ul Bitr\'an 1305, La Serena, Chile}	
	Howard A.~\textsc{Smith},\afmark{16} 
	\orcid{0009-0008-2210-4931}{Marco \textsc{Donati}},\afmark{14} 
	\orcid{0000-0002-9279-4041}{Q. Daniel \textsc{Wang}},\afmark{35} \aftext{35}{Department of Astronomy, University of Massachusetts, Amherst, MA 01003, USA} 
    and
	\orcid{0000-0002-6398-7530}{N. \textsc{Bijas}}\afmark{36} \aftext{36}{Jodrell Bank Centre for Astrophysics, School of Physics and Astronomy, The University of Manchester, Manchester M13 9PL, UK.} 
}  


\maketitle

\begin{abstract}  
Star formation activity in the Central Molecular Zone (CMZ) directly manifests itself as radio continuum free-free emission (Bremsstrahlung) and radio recombination line emission from HII regions surrounding newly formed massive stars. We derive the overall distribution of the HII regions and their fundamental properties: electron temperature ($\Te$) and emission measure ($EM$), and hence electron density in the form of two dimensional distribution maps over the CMZ by analyzing the ACES (ALMA CMZ Exploration Survey) \h40 (99.02 GHz) recombination line and 99.6 GHz continuum emission data with synthesized beam widths of $2''.45$ (0.097 pc at 8.2 kpc) and $2''.14$, respectively. We apply the 'TeEM' method ($\Te$--$EM$ mapping), which creates $\Te$ and $EM$ maps from input 2D maps of the continuum and integrated line intensity. The analysis covers the entire ACES field from $l\sim -0^\circ.6$ to $+0^\circ.8$ and from $b\sim -0^\circ.2$ to $+0^\circ.1$. The area analyzed is complete and includes previously known HII regions such as Sgr B2, Sgr B1, the Sickle, the Pistol, thermal filaments (Bridges), Sgr A HII regions, the Minispiral, and many other known HII regions. Sgr C is not included in the analysis due to the insufficient signal-to-noise ratio in the recombination line map. The mean electron temperature over the CMZ is determined to be \Tcmztext (SE:standard error of the mean, SD: pixel-to-pixel standard deviation). Some HII regions, such as Sgr B2 Main and the Minispiral, exhibit large scatter and an internal $\Te$ gradient of several thousand K per parsec. The $EM$ distribution is more diverse, varying by orders of magnitude from $\sim 10^5$ to $\sim 3\times 10^8$ \emunit within the CMZ, as well as within individual HII regions.

\end{abstract}
 
 
\section{Introduction}\label{intro}  

The central molecular zone (CMZ) of our Galaxy is a high-density molecular gas disk with a moderate star formation rate (SFR) \citep{H2023,S2022}.
One of the powerful means to measure the SFR is to observe recombination lines such as H$\alpha$ emission from HII regions embedding newly born massive stars \citep{K2012}. 
In the Galactic Center, which is opaque to optical lines, the radio recombination lines (RRL) and continuum free-free emission (Bremsstrahlung) provide efficient and direct means of determining the SFR by measuring the distribution of H II regions along with their individual physical parameters.
The ACES (ALMA CMZ Exploration Survey) has greatly expanded the possibility of investigating the detailed properties of the CMZ not only in the molecular lines but also in recombination lines and continuum emissions \citep{ACESI,ACESII,ACESIII,ACESIV,ACESV,ACESVI,ACES2024N,S2025a,S2025b,ACES2026P}.
This paper aims to clarify the overall distribution of HII regions in the CMZ and their fundamental physical properties, the electron temperature ($\Te$) and emission measure ($EM$), and hence the electron density, using the ACES line and continuum survey data.

Studies of recombination lines and continuum emission of HII regions in the Galactic Center at radio wavelengths have found their electron temperatures to be typically $\Te\sim 5000$ -- 10000 K, as obtained from the single-dish, VLA, and ALMA observations of the thermal gasses in the individual regions of Sgr B2, Sgr B1, the Sickle, the Pistol, thermal filaments (Bridges), Sgr A HII regions, and the Minispiral \citep{M1986,M1992,M1993,Z1993,P1996,L1997,L2001,T2017,T2019,Me2019,Me2022}. 
{Slightly lower temperatures have been derived in the off-plane regions of the Galactic Center Lobe  (GCL) above the CMZ with $\Te \sim  4360\pm900$ K \citep{Nagoshi+2019}.}

In this paper, we investigate the HII regions in the CMZ using the ACES observations of the \h40 line at a frequency of 99.0229 GHz with a FWHM angular resolution of $\theta_{\rm Line}=2''.45$ of the synthesized beam, corresponding to a linear resolution of $0.097$ pc at a distance of $R_0=8.178$ kpc to the GC \citep{gravity+2019}.
We use in addition the continuum data at 99.6 GHz with a bandwidth of 3.7 GHz and a synthesized beam of $\theta_{\rm Conti.}=2''.14$ \citep{ACESII,ACESV} to calculate the continuum-to-line intensity ratio, and the continuum spectral index in combination with the 93.7 and 86.6 GHz data.  
We use the fits-formatted ACES maps at the synthesized resolutions \citep{ACESI,ACESII,ACESIII} and smooth the 99.6 GHz map to the resolution of the line map when constructing the continuum-to-line ratio map. 
The detected HII regions are considered to be resolved in our $\sim 0.1$ pc beam. 

The purpose of this paper is to derive the spatial distribution ``maps'' of the electron temperature, $\Te$, and emission measure, $EM$, for the entire CMZ based on the 2D maps.
For this purpose, we apply the TeEM ($\Te$-$EM$ mapping) algorithm, which handles 2D maps and skips the sophisticated and time-consuming analysis of fitting individual line profiles.

\section{Data and method} 

\ss{Data}

We use the publicly released data products of the ALMA Central Molecular Zone Exploration Survey (ACES; project code 2021.1.00172.L, PI: S.~Longmore), a Cycle~8 ALMA Band~3 (85--102 GHz) Large Program that maps the inner $\sim100$ pc of the Galaxy over $l=359^\circ.4$ to $0^\circ.8$ and $|b|\le0^\circ.25$ \citep{ACESI}. The survey consists of 45 contiguous sub-mosaics, combining the ALMA 12-m, 7-m, and Total Power arrays — at present, the largest mosaic produced by ALMA — reaching an angular resolution of $\sim2''$, corresponding to $\sim0.1$ pc at the adopted distance. The \h40 radio recombination line (rest frequency $99.02295$ GHz) lies in one of the two broad upper-sideband spectral windows ($97.66$--$99.54$ GHz) and is released as part of the broad-window line products \citep{ACESV}, while the continuum data reduction is described by \citet{ACESII}.

For the recombination-line analysis, we use the \h40 peak-intensity and velocity-integrated (moment~0) maps, in units of Jy~beam$^{-1}$ and Jy~beam$^{-1}$~km~s$^{-1}$, respectively. The line cube has a channel width of $0.49$ MHz ($1.5$ km~s$^{-1}$ at the line frequency) in the LSRK (Kinematic Local Standard of Rest) frame, and a synthesized beam of $2''.84\times2''.11$ at ${\rm PA}=89^\circ$, i.e., a geometric-mean FWHM of $\theta_{\rm Line}=2''.45$. For the continuum, we use the ACES aggregate 3-mm image at a representative frequency of $99.6$ GHz (effective bandwidth $3.7$ GHz), restored to a circular beam of $\theta_{\rm Conti.}=2''.14$ \citep{ACESII}. All maps share a common grid of $0''.5$ pixels in Galactic coordinates (GLON--TAN, GLAT--TAN), centered near $(l,b)=(0^\circ.143,~0^\circ)$, and spanning the full ACES footprint.
The noise is spatially variable across the mosaic, increasing toward the brightest continuum sources such as Sgr~A$^*$ and Sgr~B2 \citep{ACESII}.

\begin{figure*}   
\begin{center}
\includegraphics[width=\lw]{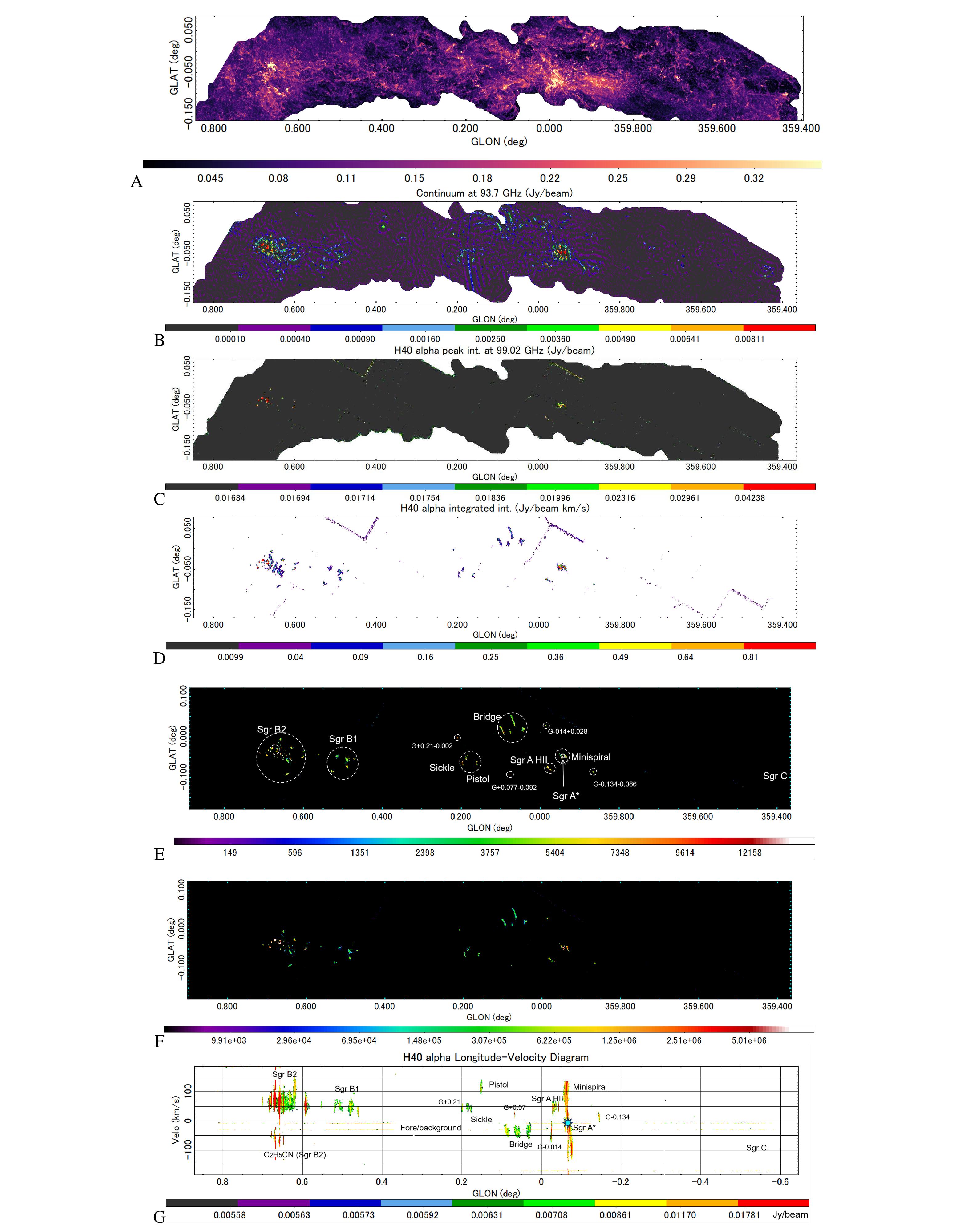}  
\end{center} 
\caption{ACES distribution maps of the CMZ.
Panel A: \cs~peak intensity map (in \Jybeam) for comparison;
Panels B, C, and D: 93.7 GHz continuum intensity (99.6 GHz map is almost identical) (\Jybeam), \h40 integrated intensity (\Jybeam \kms), and \h40 peak intensity (\Jybeam) maps, respectively;
Panels E and F: The resulting $\Te$ (in K) and $EM$ (in \emu) maps of the HII regions, respectively;
Panel G:\h40 longitude-velocity diagram (LVD) in \Jybeam.
Circles indicate representative regions. 
{Alt text: ACES maps used in the analysis and resulting $\Te$ and $EM$.}
} 
\label{fig-full}   
\end{figure*}

{Figure \ref{fig-full} shows the following maps from ACES:\\
Panel A is \cs peak intensity map, for comparison;\\
Panels B, C, and D show the 93.7 GHz continuum intensity (99.6 GHz map is very similar), \h40 integrated intensity, and \h40 peak intensity maps, respectively, taken with the same ALMA configurations;\\
Panels E and F show the resulting $\Te$ and $EM$ maps of the HII regions, respectively;\\
Panel G is the \h40 longitude-velocity diagram (LVD).}

\ss{Analysis method: TeEM algorithm}\label{ssTeEM} 

\def\MC{M$_{\rm C}$ } \def\MI{M$_{\rm LI}$ } \def\MP{M$_{\rm LP}$ } \def\MV{M$_{\delta v}$ }
\def\MRI{M$_{\rm RI}$ }\def\MRP{M$_{\rm RP}$ }\def\MET{M$_{{T}_{\rm e}}$ }\def\MEM{M$_{EM}$ }

\begin{figure}     
\begin{center}   
\includegraphics[width=\lw]{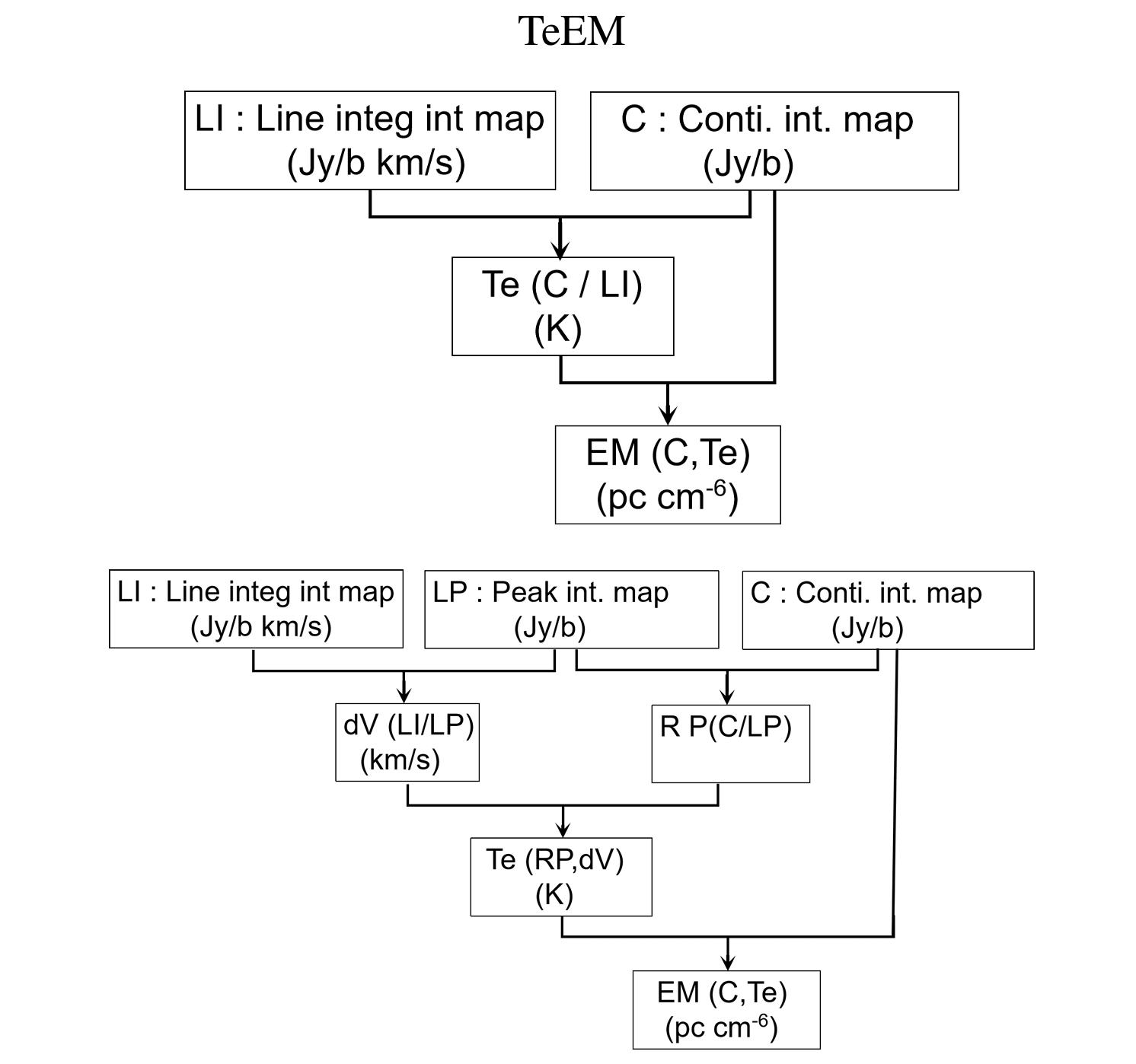}  
\end{center} 
\caption{TeEM flowchart using input 2D maps of the continuum intensity, line peak intensity and line integrated intensity to create the continuum-to-line intensity ratio, $\delta V$, $\Te$ and $EM$ maps. 
If $\Delta V$ and continuum-to-line ratio are not necessary, the top panel creates the same result.
{Alt text: TeEM flowchart. }
}
\label{fig-TeEM}  
\end{figure}

We introduce a method for creating two dimensional (2D) maps of the electron temperature, $\Te$, and emission measure, $EM$, based on the formulations described in the next subsections, using 2D fits-formatted maps. We refer to this method as ``TeEM'' ($\Te$-$EM$ mapping method).
This method makes it possible to obtain distribution maps of $\Te$ and $EM$ without losing data by skipping the handling of the individual line profiles based on 3D cubes.
The algorithm is illustrated by flowcharts in figure \ref{fig-TeEM}.

\ss{Limitation and uncertainties}

\sss{Effect of multiple velocity components}

\begin{figure}
    \begin{center} \includegraphics[width=\linewidth]{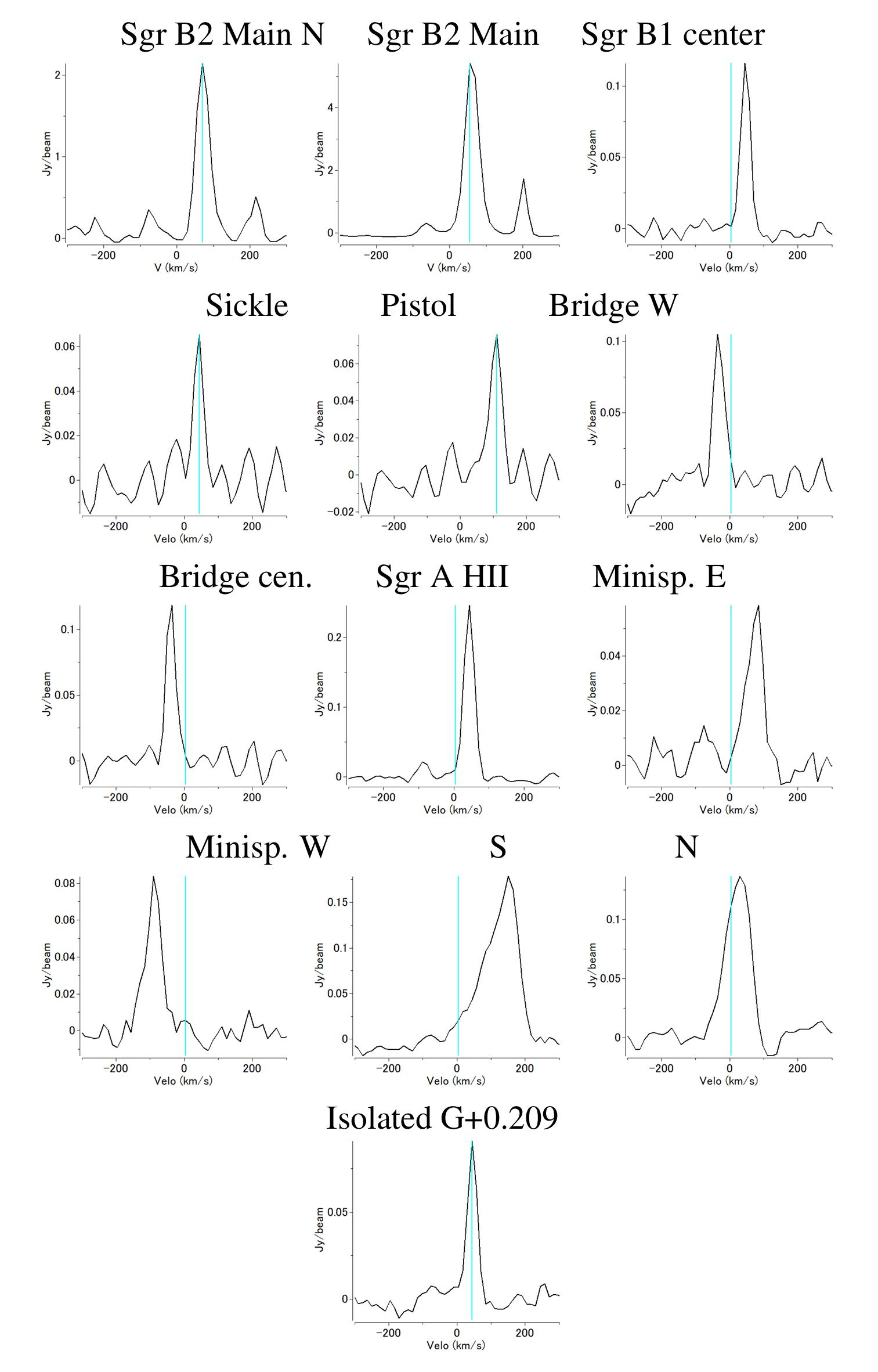} 
    \end{center}
    \caption{\h40 line profiles at representative positions of the analyzed regions, showing the profiles are relatively simple and the assumption of single Gaussian is plausible to approximate the line width by dividing  integrated intensity by peak intensity.}
    \label{fig:spec}
\end{figure}

Since the method uses 2D peak and integrated-intensity maps rather than fitting individual spectra, the derived velocity width could be affected in regions with multiple velocity components, non-Gaussian profiles, such as those with outskirts, or low signal-to-noise \h40 emission. 
This could cause an overestimation of the velocity width, resulting in an underestimation of $\Te$.
However, almost all regions show single and simple line profiles, as shown in figure \ref{fig:spec}, reflecting the fact that one source/component dominates in our small beam, we expect this effect to have a relatively small impact on the analysis.

\sss{Smaller contamination of synchrotron emission at higher frequencies: an advantage of mm-waves}

\begin{figure}   
\begin{center}     
\includegraphics[width=\lw]{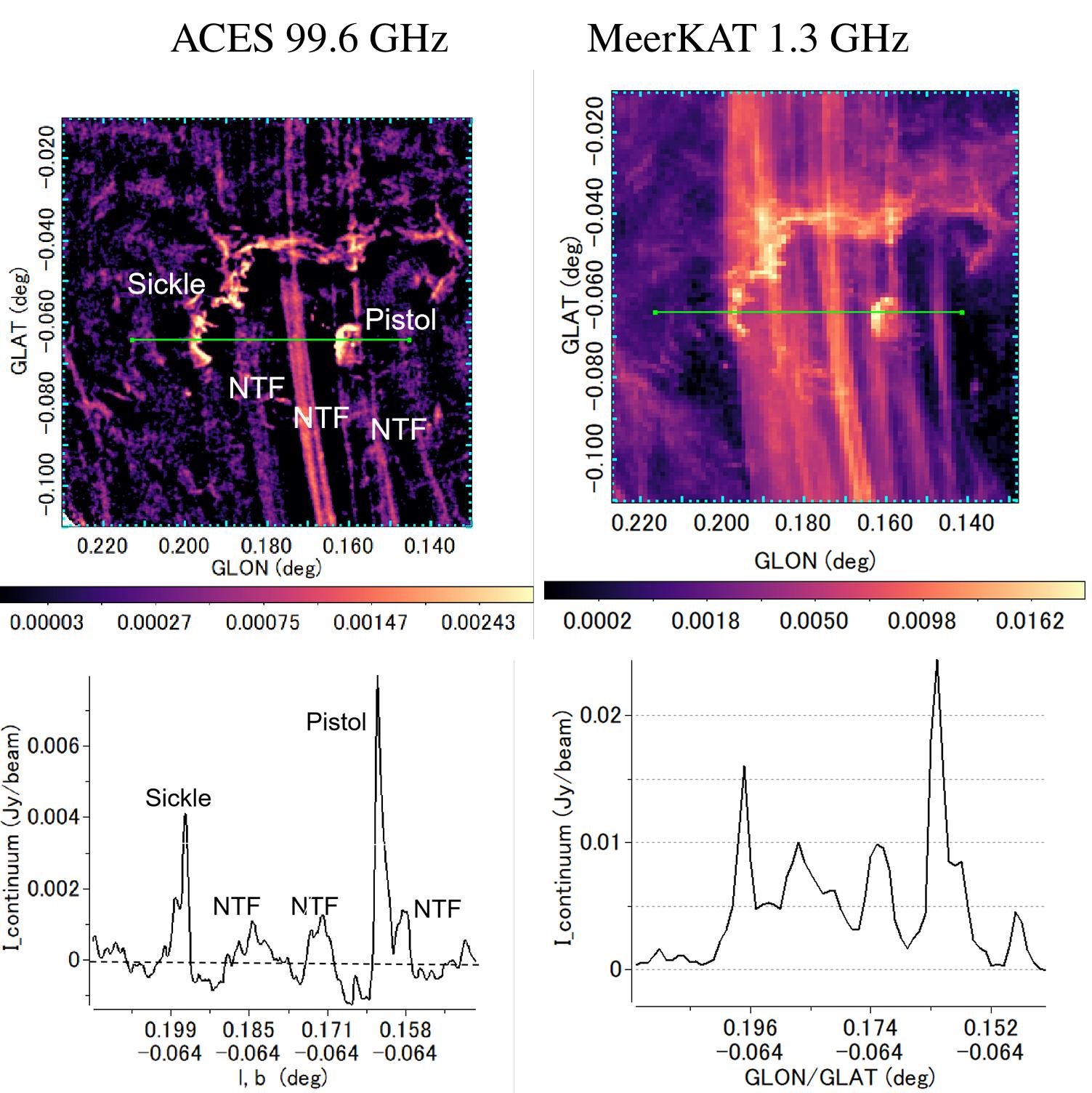}   
\end{center}  
\caption{[Top] Maps of the thermally emitting Sickle and Pistol overlapping the Radio Arc at 93.7 and 1.3 GHz \citep{H2022Mkat} continuum, demonstrating the advantage of using mm-waves over microwaves.
[Bottom] Same, but in cross section along the line inserted in the top panels.
{Alt text: Radio continuum map of the Sick and Pistol region in the Radio Arc. }
}
\label{fig-synch}  
\end{figure}

In HII regions, we may neglect the synchrotron emission at high frequencies because of the steep negative spectral index, except for some particular regions overlapping the bright non-thermal filaments (NTF) of the Radio Arc, and in HII regions penetrated by strong magnetic fields \citep{Me2019}.
In figure \ref{fig-synch} we present radio continuum maps at 99.6 GHz of the Sickle and Pistol with the Radio Arc, along with a cross section across these sources along the inserted line in the top panel, compared with those at 1.3 GHz from MeerKAT observations \citep{H2022Mkat}.
This figure demonstrates that the non-thermal emission from the filaments is an order of magnitude weaker than the free-free emission in the mm-wave range, whereas they are comparably strong in the micro-wave range.
This is an advantage of using the mm-wave data over micro-wave data in order to derive $\Te$, because the impact of subtracting synchrotron intensity can be minimized.
In other regions where the \h40 line is detected, the synchrotron emission is generally much weaker than in the NTF and may be safely neglected.
\sgrastar is the strongest non-thermal source in the analyzed field, but it is not detected in the \h40 line, hence, does not affect the present study. 

\sss{Contamination by dust continuum emission and masking}

\begin{figure*}   
\begin{center}  \includegraphics[width=\lw]{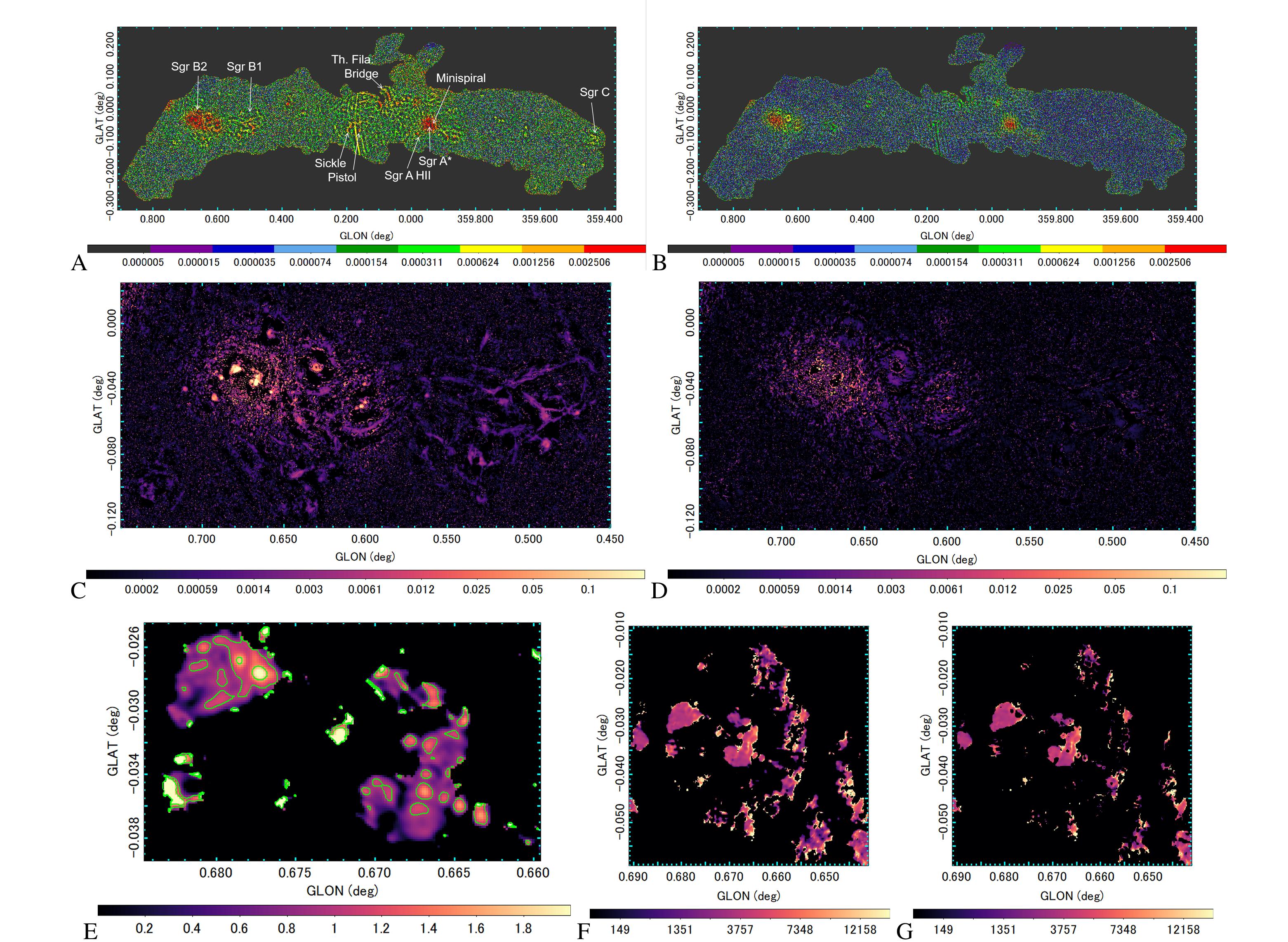}
\end{center}  
\caption{ 
[A] Free-free and [B] dust emission maps separated using 86.6 and 99.6 GHz continuum maps in order to confirm that the dust contribution is small. {[C, D] Same as panels A and B,} respectively, but enlarged for Sgr B2 and B1 regions. 
The units are J beam$^{-1}$ for all panels. 
[E] Ratio of 99.6 to 86.6 GHz continuum intensities for Sgr B2 Main. Contours are drawn at the 99.6 to 96.6 GHz intensity ratio of 0.986 and 1.631 ($\alpha=-0.1$ and $+3.5$, respectively).
[F] No masked $\Te$ of the Sgr B2 region.
[G] Same as F, but dust-dominated-regions are masked $\Te$ above $\alpha>+2.0$.
{Alt text: Free-free and dust emission maps of CMZ by ACES. }
}
\label{fig-dust-ff}  
\end{figure*}

The dust emission cannot be safely ignored in mm-wave range \citep{Sch2016,X2025}.
Therefore, we try to separate the dust contamination and minimize its contribution.
Figure \ref{fig-dust-ff} compares free-free (Bremsstrahlung) and dust emission intensities calculated at 99.6 GHz using the ACES continuum maps at 99.6 and 86.6 GHz, convolved to the 99.6 GHz beam.
We assumed spectral indices of $\alpha=-0.1$ for the free-free emission and $+3.5$ for the dust emission \citep{Sch2016,X2025}:
\be
f=f_{\rm ff}(\nu/\nu_1)^{-0.1} + f_{\rm dust} (\nu/\nu_1)^{+3.5},
\ee
where {$\nu=99.6$ GHz and $\nu_1=86.6$ GHz}.
The figure indicates that $f_{\rm ff}$ is generally brighter than $f_{\rm dust}$. 
However, there are areas with significant dust emission. 
In order to check this in more detail, we have created an intensity ratio map between 99.6 and 86.6 GHz, as shown in panel E of figure \ref{fig-dust-ff}.
The ratio is less than unity (negative $\alpha$) in most areas, but there are also knotty regions indicating contamination from dust emissions.
Thus, we made a map where extremely positive index regions with $\alpha > +2$ have been masked. Overall, the resulting $\Te$ map has not changed much; however, some differences are found in peaky strong continuum sources, such as those in Sgr B2. 
We therefore applied a mask of dust-excess regions to the entire CMZ so that pixels with a spectral index of the continuum emission exceeding $\alpha=+2$, or regions with $I_{\rm 99.6~GHz}/I_{\rm 86.6~GHz}>(99.6/86.6)^{+2}=1.323 $, are removed from the analysis.
Panels {F and G} compare the resulting $\Te$ maps of the central region of Sgr B2 before and after the dusty-region masking. 

\sss{Side lobes}

\begin{figure}     
\begin{center}   
\includegraphics[width=\lw]{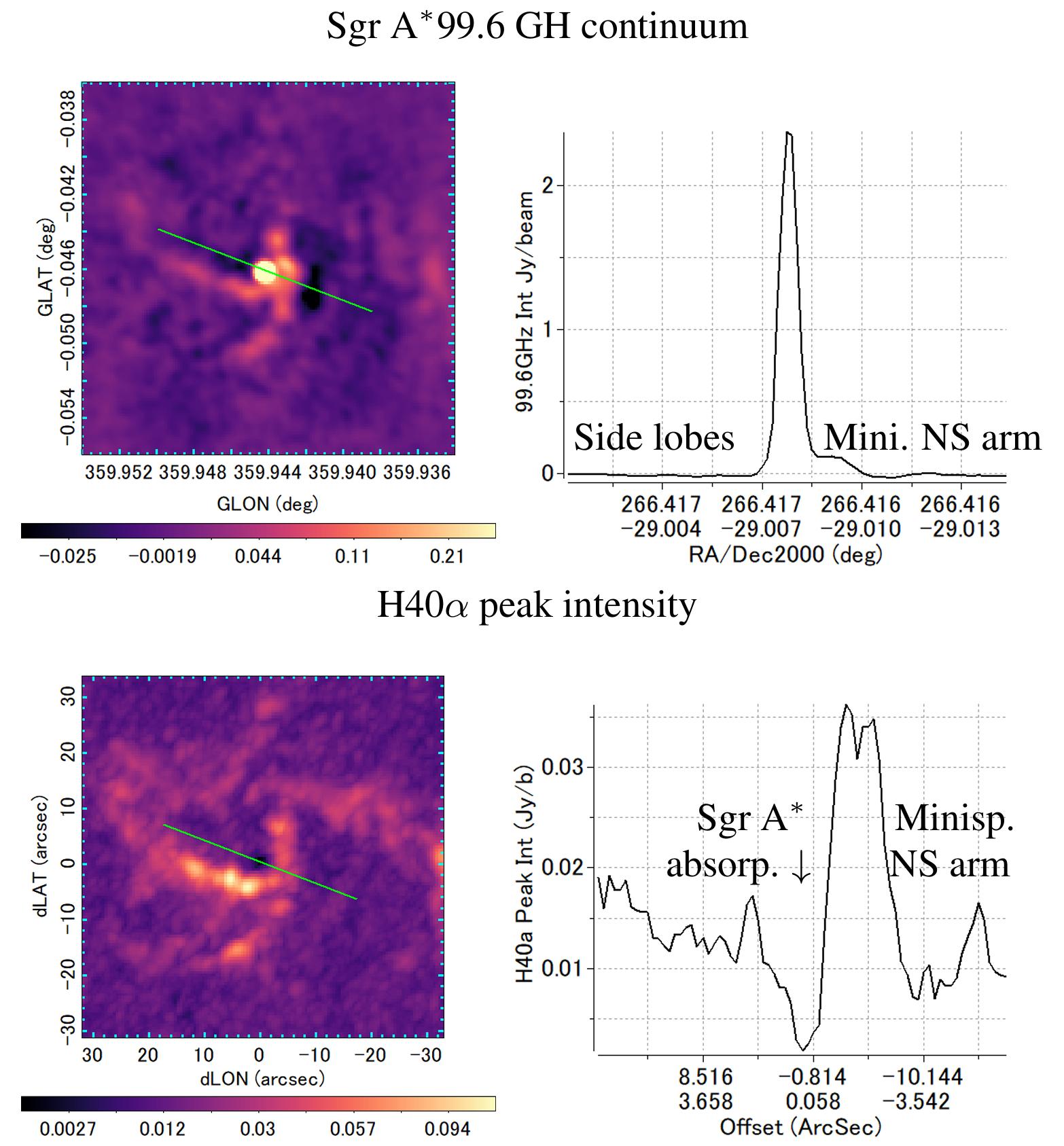}  
\end{center} 
\caption{{[Top] 99.6GHz continuum around \star and a cross section along the inserted line through a 'quiet' region, showing that the side lobe level is $\lesssim 1$\% of the main beam center intensity.
[Bottom] Same in \h40-peak intensity. \star appears as a sharp absorption.
{Alt text: 99.6 GHz side lobe and . }}
}
\label{fig:sidelobe}  
\end{figure}

Another factor that potentially impacts the analysis is the side lobes around strong radio sources such as \star and Sgr B2 Main. 
Figure \ref{fig:sidelobe} shows a 99.6GHz continuum map near \star and a cross section across the central source.
The side lobe level is less than $\sim 1$ percent of the main beam center intensity, which is generally much less than the \h40 line intensities in the CMZ, except for the neighboring regions of such exceptionally strong emission sources like \star.

\star is the strongest point-like source in the continuum, but it does not emit the \h40 line; hence, it does not appear in the $\Te$ map but appears in the \h40 map as an absorbing point source.
Its continuum side lobes may contaminate the surrounding continuum emission, but they do not necessarily overlap with the HII regions. 
A similar situation occurs around Sgr B2 Main, which is an emitter of both the continuum and the \h40 line.

On the other hand, the side lobes in the \h40 line are sufficiently lower than the clipping level used to create the moment 0 (integrated intensity) maps, so they do not affect the results. 

So, we consider that the side lobe effects are generally small and do not disturb the mapping of the entire CMZ and the statistics. However, we must be careful to discuss detailed properties closest to the strong continuum sources.

\sss{Contamination by molecular lines}
 
Contamination of molecular lines at rest frequencies of 99.118737, 99.069191, and 98.976705 \citep{2013A&A...559A..47B}, corresponding to +210, -70, and -220 \kms in our \h40 spectra, causes an overestimation of the line intensity if the molecular cloud overlaps in the line of sight.
Such a rare case is found in Sgr B2 Main as shown in  figure \ref{fig:spec}.   
However, the peak intensities of the contaminated molecular lines are an order of magnitude weaker than the \h40 peak intensity, and the line width is much narrower.
Thus, the effect of such contamination is on the order of $\sim 10$\% in $\Te$, and affects only a relatively small area of Sgr B2 Main.

Except for Sgr B2 Main, we do not see any line contamination, including the larger area surrounding Sgr B2 Main, as shown in figure \ref{fig:spec}, insofar as the currently used maps are concerned.
This is more clearly demonstrated by the longitude-velocity diagram shown in figure \ref{fig-full}-D.
Therefore, we are neglecting the effect of line contamination in the analysis.

\sss{Edging effect of the line intensity (moment 0) map}

The moment 0 (integrated intensity) map of the \h40 line emission covers only the regions where the intensity is greater than the threshold or several times the rms noise level of the spectral line maps (3D cube).
This clipping causes a sharp cutoff in the integrated intensity of the \h40 line emission around HII regions, which directly propagates to the $\Te$ map, causing an artificially sharp edge surrounding the HII region. 
However,this does not affect the calculation of $\Te$ and $EM$.

\subsection{Electron temperature}

\def\TL{T_{\rm L}} \def\TC{T_{\rm C}} \def\IP{I_{\rm L:peak}} \def\IL{I_{\rm L}}

The electron temperature of ionized gas in local thermal equilibrium (LTE) in an HII region can be derived by measuring the intensity of the radio continuum and recombination line emission \citep{Mezger+1967II,L1976,S1983,Q2006,W2015,T2017}.
We use the expression written in terms of the brightness temperatures of the continuum and line emissions, $\TC$ and $\TL$, respectively, as given by \citet{Q2006};
\be
{\Te \over {\rm K}}=\left[7103.3 
\left(\nu_{\rm L} \over {\rm GHz} \right)^{1.1}
\left(\frac{\TC}{\TL} \right)
\left(\Delta v \over {\rm km~s^{-1}}\right)^{-1}  \left(1+{{\rm [He]}\over{\rm [H]}}\right)^{-1}
\right]^{0.87}. 
\label{eqTe}
\ee
When the continuum-to-line intensity ratio is calculated, we fix the frequency to that of the \h40 line, $\nu=\nu_{\rm L}=99.022950$~GHz.
The above equation can be re-written in terms of the continuum intensity $\IC ~[\eJybeam]$ and integrated line intensity $\IL~[\eJybeam~\ekms]$ as \citep{Rohlfs+2000,T2017}
\be
{\Te \over {\rm K}}=\left[{6985 \over a(\nu,\Te)} 
\left(\nu_{\rm L} \over {\rm GHz} \right)^{1.1}
\left(\frac{\IC}{\IL} \right)
\left(1+{{\rm [He]}\over{\rm [H]}}\right)^{-1}
\right]^{1/1.15},
\label{eq:TeTsu}
\ee
where $a(99 {\rm GHz}, \sim 8000~{\rm K})\simeq 0.9$ is a correction factor close to unity \citep{Mezger+1967I}.
We assume [He]/[H]=0.07, so the equation reduces to   
\be 
{\Te\over {\rm K}} \simeq 1.845\times 10^5  \left[ \IC (\nu_{\rm L})\over \IL \right]^{1/1.15}.
\label{eq:TeF}
\ee       
We also assume that the line profiles are relatively simple, as observed in the local HII regions \citep{M1992,M1993,L1997,L2001}.
Nevertheless, contamination by a background or foreground HII region could affect the intensity as well as the velocity width, and in such cases, analysis of the line profiles from a 3D cube is necessary.
However, since overlapping is considered rare, as suggested by the line profiles shown in figure \ref{fig:spec} and the present paper aims to study global properties, we ignore such multiple line effects here.  

\subsection{Emission measure }

Given the electron temperature $\Te$, we can calculate the emission measure $EM$ \citep{O1961} using the continuum map at any frequency (here 99.6 GHz). 
The brightness temperature $\TC$ of the continuum emission from an optically thin plasma is related to the electron temperature and optical depth $\tau$ as
\be
\TC={\lambda^2 \over 2k} \IC\simeq \tau \Te,
\ee 
and the optical depth $\tau$, which is related to the emission measure \citep{O1961}, has been re-written as follows for radio astronomy applications \citep{Mezger+1967I,Mezger+1967II}: 
\be
\tau=3.014\times 10^{-2}~\A~ \TK^{-3/2}\nuG^{-2} EM
\label{eq:tau}
\ee
where  
\be 
\A=\ln\left[4.955\times 10^{-2} \nuG^{-1} \TK ^{3/2} \right]
\ee
is a slow function of $\nu$ and $\Te$ according to the integral of the Coulomb cross section.
At $\nu\sim 99$ GHz and $\Te\sim 8000$ K, we have $\A\simeq 5.87$.
 
We thus obtain a relation to calculate $EM$ in terms of $\Te$ and $\IC$ at $\nu$ as 
\be
\left(EM\over {\rm pc~cm^{-6}} \right)
= 869.6 ~\A^{-1}\left( \IC\over {\rm Jy ~beam^{-1}}\right)
\left(\nu \over{\rm GHz} \right)^2
\left(\Te \over {\rm K} \right)^{1/2}\label{eqEMoster} 
\ee  

\ss{Electron density}
    
The mean electron density $\Ne$ over the line of sight is then obtained as
\be
\left<\Ne \right> \sim \sqrt{EM/L} \label{eq:ne}
\ee
where $L$ is the line-of-sight length of the region.
In our data, we have no means to measure $L$ directly; however, an order of magnitude estimate is possible by assuming that $L$ is comparable to the projected width or diameter of the region. 

\ss{Example of analysis for the Pistol}
 
\begin{figure*}   
\begin{center}  
\includegraphics[width=\lw]{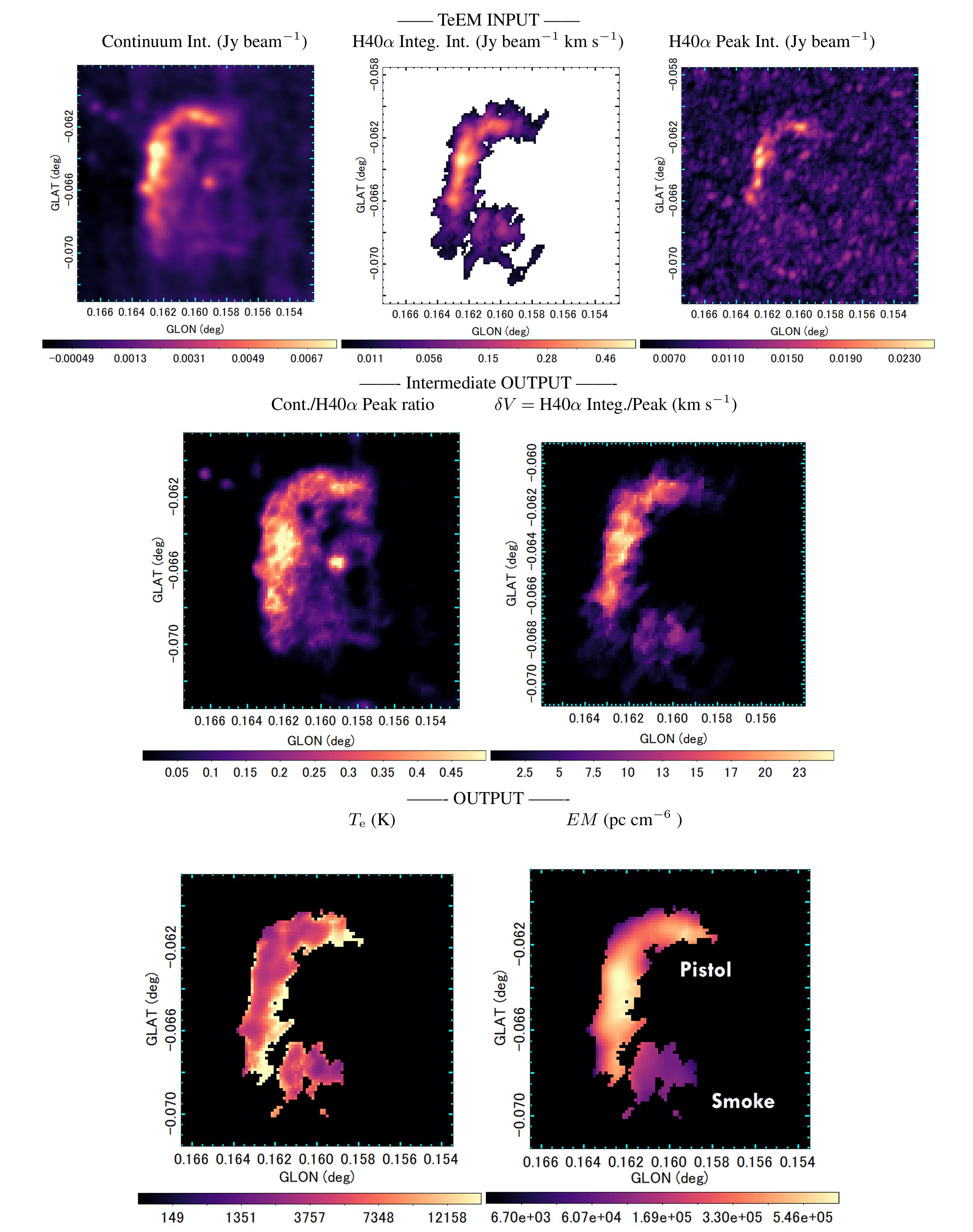}
\end{center} 
\caption{
[Top] Input 2D maps for TeEM: (left) the continuum intensity, (middle) \h40 line integrated intensity, and (right) peak intensity of the Pistol region.
[Middle row] (left) Continuum-to-line intensity ratio, (right) Velocity width.
[Bottom] Resulting maps by TeEM: (left) $\Te$ and (right) $EM$ maps.
Here and hereafter, the color scalings for $\Te$ is valid only above $\sim 1000$ K.
{Alt text: Input and output 2D maps by TeEM for Pistol region.}
}
\label{fig-pistol}  
\end{figure*}

Applying the TeEM method to the fits-formatted two dimensional maps of the continuum emission at 99.6 GHz, \h40 integrated and peak intensities at 99.02 GHz, we obtained $\Te$ and $EM$ maps for HII regions in the CMZ covered by ACES. 
As an example, we present the result for the Pistol (G+0.16-0.06), as shown in figure~\ref{fig-pistol}. The first three panels show the input 2D maps of continuum, integrated intensity, and peak intensity of the \h40 emission. 
The second row panels are the calculated results {for the continuum to line ratio and  $\delta V$,} and the third row shows the results for $\Te$ and $EM$: 
$\Te$ of the Pistol varies from $\sim 5000$ to 7000 K from north to south, while $EM$ has a mild maximum near the midpoint at $\sim 7\times 10^5$ \emu.
The ``smoke'' in the south is a faint nebula with low $EM$ of $\sim 10^5$ \emu, but the temperature almost the same as the remaining part of the region, and uniform at $\sim 5000$-- 7000 K.  

\def\ghz{{\rm GHz}}

\section{Electron temperature and emission measure}\label{sec:TeEM}

\ss{The maps }

\begin{figure*}   
\begin{center}       
\includegraphics[width=\lw]{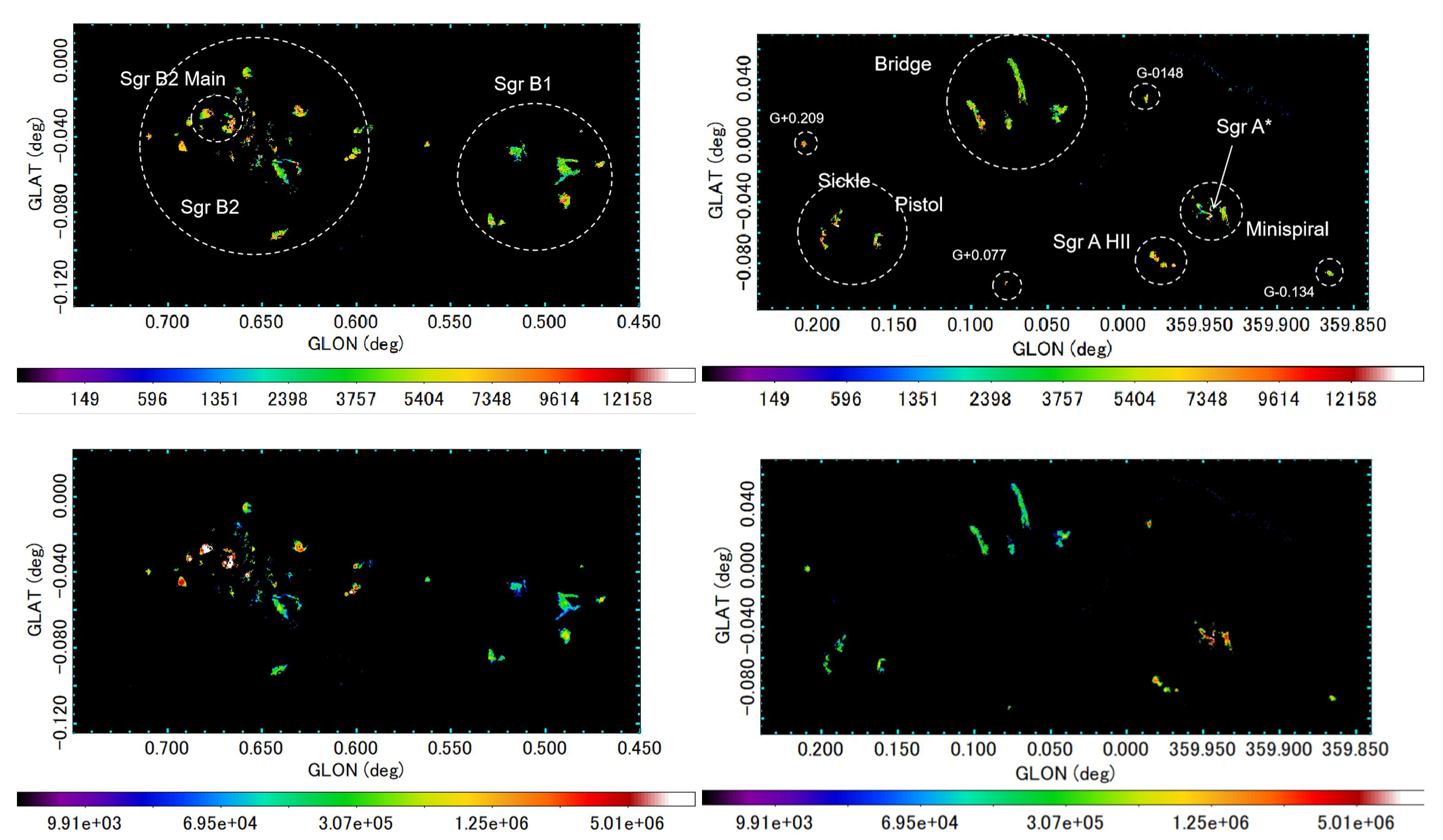}    
\end{center} 
\caption{Same as bottom two panels of figure \ref{fig-full}, but enlarged. $\Te$ (top and 3rd) and $EM$ (2nd and bottom) maps. Circles indicate representative regions.
{Alt text: $\Te$ maps.}
}
\label{fig-semiful} 
\end{figure*}

Figure \ref{fig-full} shows the maps used in the analysis: (A) the intensity of radio continuum emission at 99.6 GHz, (B) the peak intensity, and (C) the integrated intensity of the \h40 line, along with the (D) LVD of the \h40 line.  
The resulting distribution maps of the electron temperature $\Te$ (K) and emission measure $EM$ (\emunit) are shown in the bottom two panels (E and F) of figure \ref{fig-full}, respectively, and are enlarged in figure \ref{fig-semiful}. 

\subsection{General properties}

Generally, the \h40 line emitting regions are clustered together with star forming regions such as Sgr B2 and B1, the Sgr A HII regions, the thermal Bridge, and the Minispiral.
There are several isolated compact HII regions that appear to be without connections to extended HII features.
Sgr C is not visible in the electron temperature and emission measure maps due to its very weak integrated intensity in the \h40 line.
{Such a localized distribution of HII regions can be attributed to the interferometer observations at 99 GHz insensitive to diffuse emissions both in continuum and \h40 line as well as to the finite clipping level to produce the integrated line intensity map.}

\ss{Averaged $\Te$ and table of the result}

Although we presented the $\Te$ values in the form of maps, presenting typical values will be useful for discussing the physics of the HII regions in the CMZ.
So, we calculate the mean value $\Tav$ as an average of $\Te$ for all $N$ pixel points with meaningful calculated values in a desired area of the map (bin area):
$\Tav=\Sigma \Te / N$.
The errors are given by the pixel-to-pixel standard deviation (SD) and standard error (SE) defined by 
$\sigma_{\rm SD}=\sqrt{\Sigma(\Te-\Tav)^2/N}$  and  
$\sigma_{\rm SE}=\sigma_{\rm SD}/\sqrt{N_{\rm data}}$ , respectively.
Here  
$N_{\rm data}\sim N/({\rm size /beam ~width})^2 $ is the number of independent observed data points.

We first calculate a grand average $\Tcmz=\Tav_{\rm Entire~CMZ}$ of the obtained $\Te$ values in the whole CMZ shown in figure \ref{fig-full}, and obtain 
\Tcmztext.

\begin{figure*}   
\begin{center}  
\includegraphics[width=\lw]{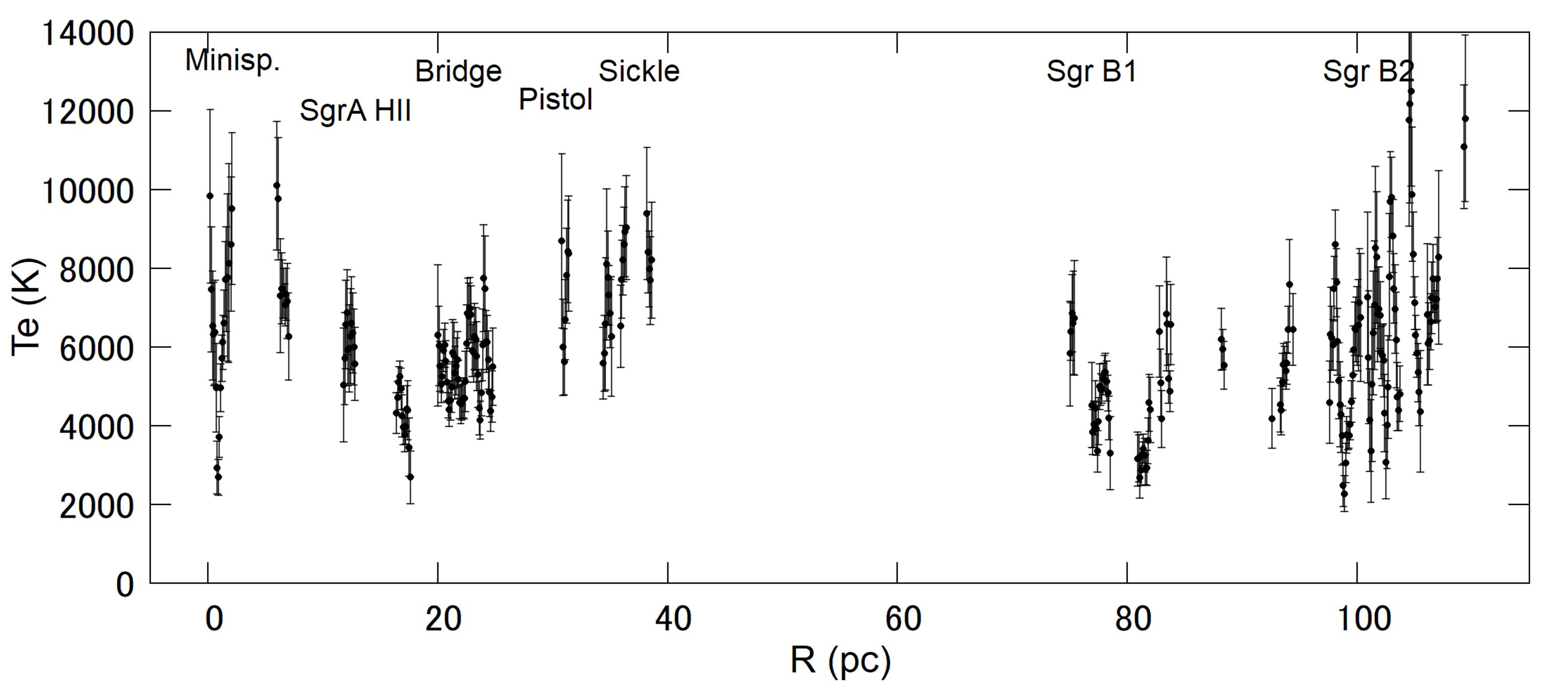}  
\end{center} 
\caption{Radial variation of averaged $\Te$ of HII regions in the CMZ as a function of the projected distance $R$ from \sgrastar, as measured using figure \ref{fig-full}.
Running averaged $\Te$ and the standard errors by radial binning with interval and width of $\Delta R=0.1$ pc and pixel number greater than 100 per bin are plotted. 
{Alt text: Electron temperature as a function of the Galacto-centric distance.}}
\label{fig:te-r}  
\end{figure*}

Then we calculate $\Tav$ as a function of the projected distance $R$ (galacto-centric radius) from \star in radial bins with a width and interval of $\delta R=0.1$ pc. 
Thus calculated $\Tav$ is listed in table \ref{tab:TeCMZ} and plotted in figure \ref{fig:te-r} against $R$. 

Electron temperatures are generally uniform over the CMZ within $\Te\sim 4000$ to $\sim8000$ K, yielding the mean value as above. 

The emission measure is more diverse, ranging from $EM\sim 10^5$ to $\sim 3\times 10^8$ \emunit, which is roughly proportional to the continuum intensity.

\begin{table*}[]
    \caption{Radius means of $\Te$ between $R$ and $R+\delta R$ (projected distance from \star) every 0.1 pc and width $\delta R =0.1$ pc. Radius bins without emission are skipped. The values are plotted in figure \ref{fig:te-r}. The grand mean in the entire CMZ is \Tcmztext.
} 
\label{tab:TeCMZ}
    \centering
    \begin{tabular}{ccc|ccc|ccc|ccc|ccc}
    \hline
    \hline
$R$  &$\Te$ &$\pm {\rm SE}$ &$R$  &$\Te$ &$\pm {\rm SE}$ &$R$  &$\Te$ &$\pm {\rm SE}$ &R &$\Te$ &$\pm {\rm SE}$  &R &$\Te$ &$\pm {\rm SE}$ \\  
(pc)&(K) &(K)&(pc)&(K) &(K)&(pc)&(K) &(K)&(pc)&(K) &(K)&(pc)&(K) &(K)\\  
 \hline 
      0.2 &  9834. &  2206.&     0.3 &  7464. &  1588.&     0.4 &  6545. &  1391.&     0.5 &  6331. &  1305.&     0.6 &  6376. &  1318.\\
      0.7 &  4968. &  1125.&     0.8 &  2939. &   676.&     0.9 &  2694. &   458.&     1.0 &  3717. &   518.&     1.1 &  4966. &   609.\\
      1.2 &  5714. &   584.&     1.3 &  6119. &   692.&     1.4 &  6605. &   843.&     1.5 &  7721. &   951.&     1.6 &  7725. &  1326.\\
      1.7 &  7763. &  2126.&     1.8 &  8131. &  2525.&     2.0 &  8616. &  1702.&     2.1 &  9515. &  1926.&     6.0 & 10100. &  1639.\\
      6.1 &  9770. &  1550.&     6.3 &  7306. &  1449.&     6.4 &  7488. &   900.&     6.5 &  7479. &   738.&     6.6 &  7372. &   636.\\
      6.7 &  7078. &   535.&     6.8 &  7339. &   660.&     6.9 &  7169. &   958.&     7.0 &  6270. &  1108.&    11.8 &  5039. &  1447.\\
     11.9 &  5710. &  1174.&    12.0 &  6573. &  1127.&    12.1 &  6883. &  1076.&    12.2 &  5931. &   904.&    12.3 &  5969. &  1111.\\
     12.4 &  6277. &  1207.&    12.5 &  6614. &  1178.&    12.6 &  6366. &  1018.&    12.7 &  5999. &   966.&    12.8 &  5570. &   934.\\
     16.4 &  4326. &   517.&    16.5 &  4715. &   406.&    16.6 &  5111. &   382.&    16.7 &  5245. &   399.&    16.8 &  4941. &   530.\\
     16.9 &  4246. &   534.&    17.0 &  3967. &   450.&    17.1 &  3771. &   433.&    17.2 &  3979. &   443.&    17.3 &  4435. &   582.\\
     17.4 &  4397. &   748.&    17.5 &  3454. &   739.&    17.6 &  2691. &   673.&    20.0 &  6297. &  1797.&    20.1 &  6028. &  1014.\\
     20.2 &  5515. &   638.&    20.3 &  5062. &   474.&    20.4 &  5249. &   415.&    20.5 &  5912. &   531.&    20.6 &  6049. &   562.\\
     20.7 &  5651. &   516.&    20.8 &  5105. &   476.&    20.9 &  4619. &   428.&    21.0 &  4413. &   431.&    21.1 &  4638. &   501.\\
     21.2 &  4987. &   646.&    21.3 &  5852. &   842.&    21.4 &  5788. &   843.&    21.5 &  5343. &   671.&    21.6 &  5514. &   725.\\
     21.7 &  5686. &   710.&    21.8 &  5179. &   479.&    21.9 &  4599. &   426.&    22.0 &  4558. &   502.&    22.1 &  4652. &   501.\\
     22.2 &  4703. &   509.&    22.3 &  4696. &   518.&    22.4 &  5122. &   765.&    22.5 &  6094. &   982.&    22.6 &  6860. &   890.\\
     22.7 &  6794. &   834.&    22.8 &  6983. &   782.&    22.9 &  6826. &   722.&    23.0 &  5907. &   657.&    23.1 &  5865. &   755.\\
     23.2 &  6269. &   833.&    23.3 &  6206. &   802.&    23.4 &  5769. &   873.&    23.5 &  5306. &   847.&    23.6 &  4446. &   684.\\
     23.7 &  4141. &   487.&    23.8 &  4840. &   597.&    23.9 &  6054. &   911.&    24.0 &  7752. &  1355.&    24.1 &  7482. &  1333.\\
     24.2 &  6144. &   784.&    24.3 &  6119. &   689.&    24.4 &  5677. &   744.&    24.5 &  4851. &   614.&    24.6 &  4374. &   515.\\
     24.7 &  4741. &   656.&    24.8 &  5507. &   982.&    30.8 &  8694. &  2215.&    30.9 &  5999. &  1223.&    31.0 &  5620. &   840.\\
     31.1 &  6696. &  1023.&    31.2 &  7817. &  1202.&    31.3 &  8431. &  1307.&    31.4 &  8377. &  1465.&    34.4 &  5596. &   910.\\
     34.5 &  5838. &   966.&    34.6 &  6594. &  1676.&    34.7 &  8100. &  1920.&    34.8 &  7764. &  1189.&    34.9 &  7314. &   757.\\
     35.0 &  6853. &   868.&    35.1 &  6262. &  1519.&    35.9 &  6530. &  1048.&    36.0 &  7721. &   985.&    36.1 &  8209. &   886.\\
     36.2 &  8612. &   948.&    36.3 &  8931. &  1138.&    36.4 &  9032. &  1324.&    38.2 &  9397. &  1675.&    38.3 &  8410. &  1035.\\
     38.4 &  7980. &   967.&    38.5 &  7692. &  1116.&    38.6 &  8206. &  1467.&    75.0 &  5836. &  1327.&    75.1 &  6392. &   727.\\
     75.2 &  6856. &   984.&    75.3 &  6615. &  1315.&    75.4 &  6741. &  1448.&    76.9 &  4520. &  1080.&    77.0 &  3835. &   574.\\
     77.1 &  4037. &   583.&    77.2 &  4455. &   687.&    77.3 &  3933. &   679.&    77.4 &  3354. &   531.&    77.5 &  4113. &   601.\\
     77.6 &  4993. &   572.&    77.7 &  4919. &   396.&    77.8 &  4934. &   352.&    77.9 &  5192. &   402.&    78.0 &  5367. &   422.\\
     78.1 &  5356. &   477.&    78.2 &  5133. &   506.&    78.3 &  4832. &   456.&    78.4 &  4200. &   548.&    78.5 &  3312. &   927.\\
     80.9 &  3156. &   686.&    81.0 &  3175. &   594.&    81.1 &  2681. &   520.&    81.2 &  2879. &   398.&    81.3 &  3294. &   305.\\
     81.4 &  3421. &   359.&    81.5 &  3253. &   434.&    81.6 &  2886. &   386.&    81.7 &  2924. &   444.&    81.8 &  3623. &   580.\\
     81.9 &  4585. &   727.&    82.0 &  4405. &   824.&    82.8 &  6395. &  1162.&    82.9 &  5088. &   857.&    83.0 &  4176. &   726.\\
     83.4 &  6846. &  1447.&    83.5 &  6597. &  1059.&    83.6 &  5190. &   625.&    83.7 &  4874. &   522.&    83.8 &  6572. &  1011.\\
     88.2 &  6203. &   786.&    88.3 &  5940. &   509.&    88.4 &  5536. &   611.&    92.6 &  4188. &   753.&    93.3 &  4533. &   688.\\
     93.4 &  4399. &   629.&    93.5 &  5115. &   721.&    93.6 &  5546. &   546.&    93.7 &  5588. &   432.&    93.8 &  5394. &   463.\\
     93.9 &  5598. &   536.&    94.0 &  6441. &   595.&    94.1 &  7590. &  1134.&    94.4 &  6452. &   914.&    97.6 &  4598. &  1039.\\
     97.7 &  6320. &  1203.&    97.8 &  6227. &   811.&    97.9 &  6054. &   648.&    98.0 &  7481. &   822.&    98.1 &  8606. &   880.\\
     98.2 &  7638. &   863.&    98.3 &  6139. &   848.&    98.4 &  5141. &   960.&    98.5 &  4546. &  1087.&    98.6 &  4281. &   953.\\
     98.7 &  3751. &   752.&    98.8 &  2493. &   551.&    98.9 &  2276. &   452.&    99.0 &  3051. &   498.&    99.1 &  3769. &   464.\\
     99.2 &  3748. &   361.&    99.3 &  3745. &   319.&    99.4 &  4041. &   405.&    99.5 &  4614. &   532.&    99.6 &  5285. &   584.\\
     99.7 &  5922. &   613.&    99.8 &  6466. &   683.&    99.9 &  6449. &   819.&   100.0 &  6534. &  1004.&   100.1 &  6551. &  1156.\\
    100.2 &  7132. &  1362.&   100.3 &  6756. &  1626.&   100.9 &  7261. &  2174.&   101.0 &  5739. &  1683.&   101.1 &  4141. &  1301.\\
    101.2 &  3363. &  1308.&   101.3 &  5058. &  1958.&   101.4 &  6358. &  1572.&   101.5 &  7067. &  1634.&   101.6 &  8524. &  2063.\\
    101.7 &  8278. &  1669.&   101.8 &  6833. &  1200.&   101.9 &  6965. &  1067.&   102.0 &  6812. &   850.&   102.1 &  5859. &   664.\\
    102.2 &  5783. &   772.&   102.3 &  5662. &   922.&   102.4 &  4322. &   979.&   102.5 &  3081. &   936.&   102.6 &  4025. &  1119.\\
    102.7 &  4986. &  1309.&   102.8 &  7788. &  1608.&   102.9 &  9700. &  1270.&   103.0 &  9802. &  1014.&   103.1 &  8823. &   932.\\
    103.2 &  7485. &   885.&   103.3 &  6959. &  1122.&   103.4 &  6176. &  1216.&   103.5 &  4741. &   868.&   103.6 &  4395. &   518.\\
    103.7 &  4812. &   699.&   104.5 & 11770. &  2690.&   104.6 & 12180. &  2512.&   104.7 & 12500. &  2407.&   104.8 &  9884. &  1701.\\
    104.9 &  8352. &  1079.&   105.0 &  7117. &   675.&   105.1 &  6298. &   501.&   105.2 &  5846. &   608.&   105.3 &  5350. &   734.\\
    105.4 &  4854. &   857.&   105.5 &  4363. &  1542.&   106.1 &  6831. &  1792.&   106.2 &  6083. &  1041.&   106.3 &  6157. &   731.\\
    106.4 &  6639. &   716.&   106.5 &  7252. &   917.&   106.6 &  7738. &   864.&   106.7 &  7208. &   572.&   106.8 &  7013. &   409.\\
    106.9 &  7221. &   569.&   107.0 &  7734. &  1060.&   107.1 &  8283. &  2204.&   109.3 & 11090. &  1570.&   109.4 & 11810. &  2121.\\
 \hline
\hline 
\end{tabular}
\end{table*}

\ss{Extrem
ely asymmetric distribution of HII regions}

Insofar as the $\Te$ and $EM$ maps in figure \ref{fig-full} produced by using the integrated and peak intensity maps of \h40 line are concerned, the distribution of HII regions in the CMZ is extremely asymmetric about \sgrastar.
The asymmetry is most clearly demonstrated in the LVD in figure \ref{fig-full}-D.
No significant HII region, and thus no SF region, is present at the negative longitude side of the GC, except for Sgr C.
Sgr C is obviously a well known SF region on the western side of the CMZ, but it is not seen in the data used in the present analysis.

\section{Individual regions}

\begin{figure*}   
\begin{center}     
\includegraphics[width=\lw]{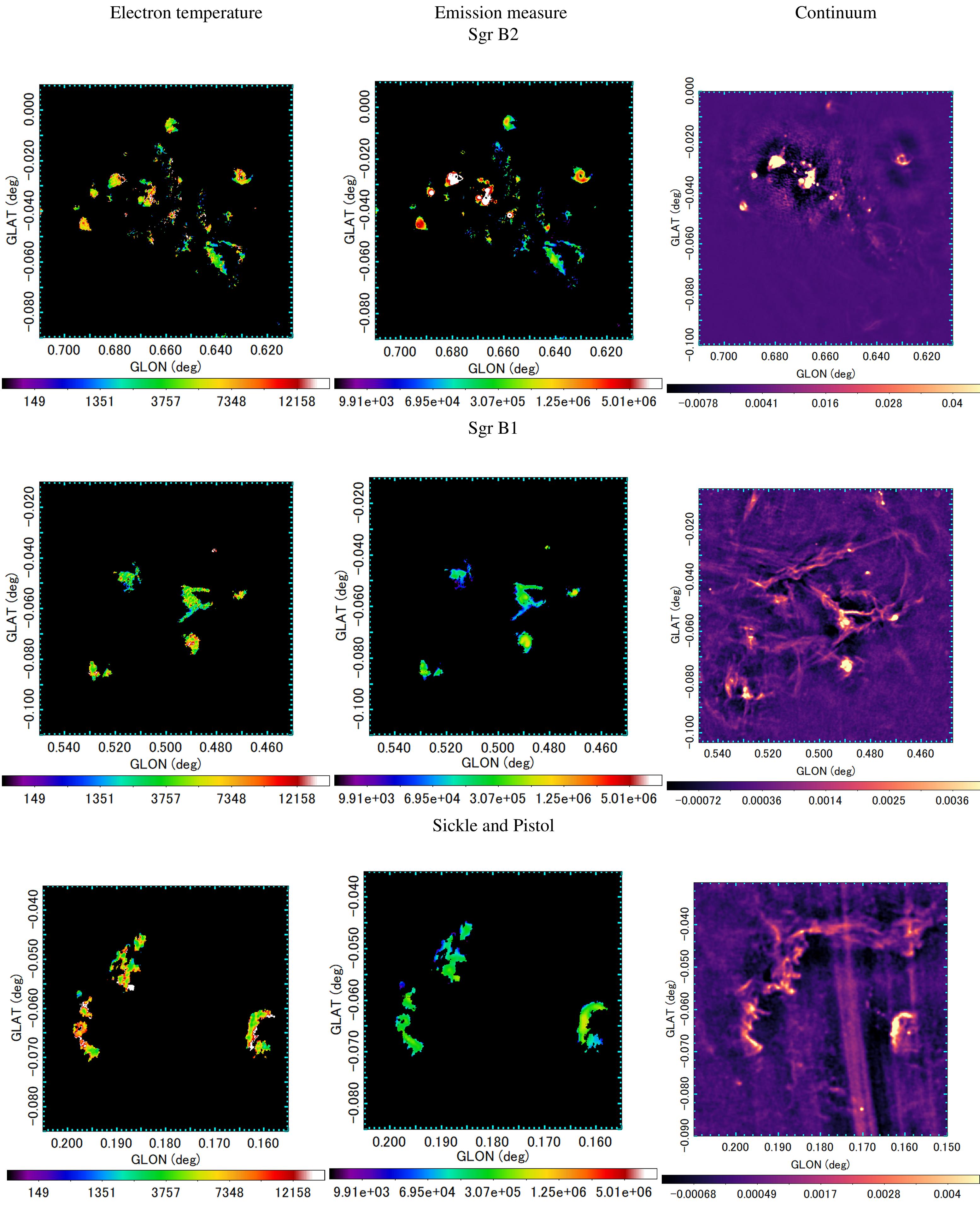}
\end{center} 
\caption{Electron temperature, emission measure and continuum maps of the individual regions in the CMZ.
{Alt text: Electron temperature, emission measure and continuum maps of the individual regions in the CMZ.}
}
\label{fig_indiv1}  
\end{figure*}

\begin{figure*}    
\begin{center}
\includegraphics[width=\lw]{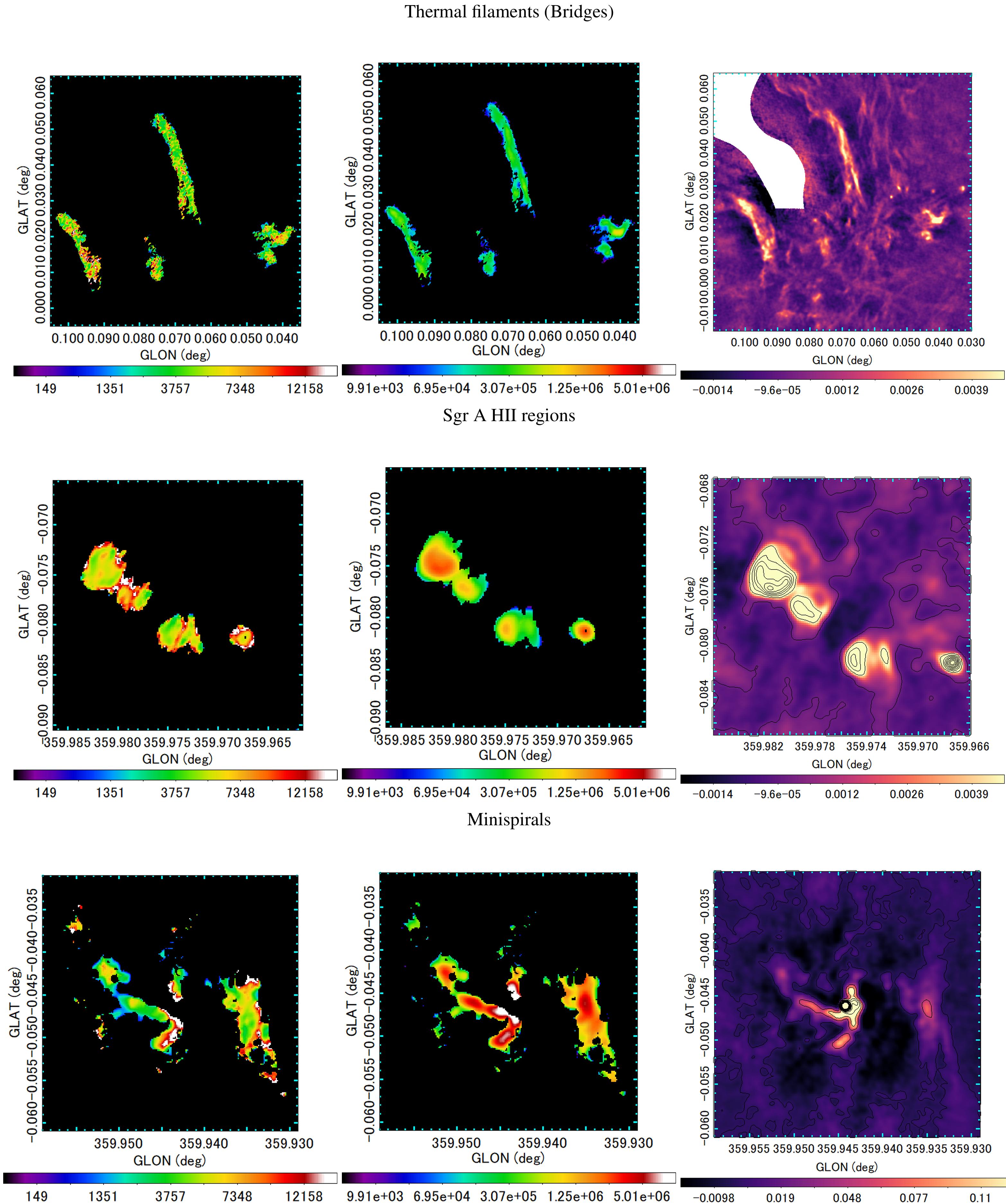} 
\end{center} 
\caption{Same as figure \ref{fig_indiv1}, but for other well known sources.
{Alt text: Same as figure \ref{fig_indiv1}.}
}
\label{fig_indiv2}  
\end{figure*}

\begin{figure*}   
\begin{center}          
\includegraphics[width=.93\lw]{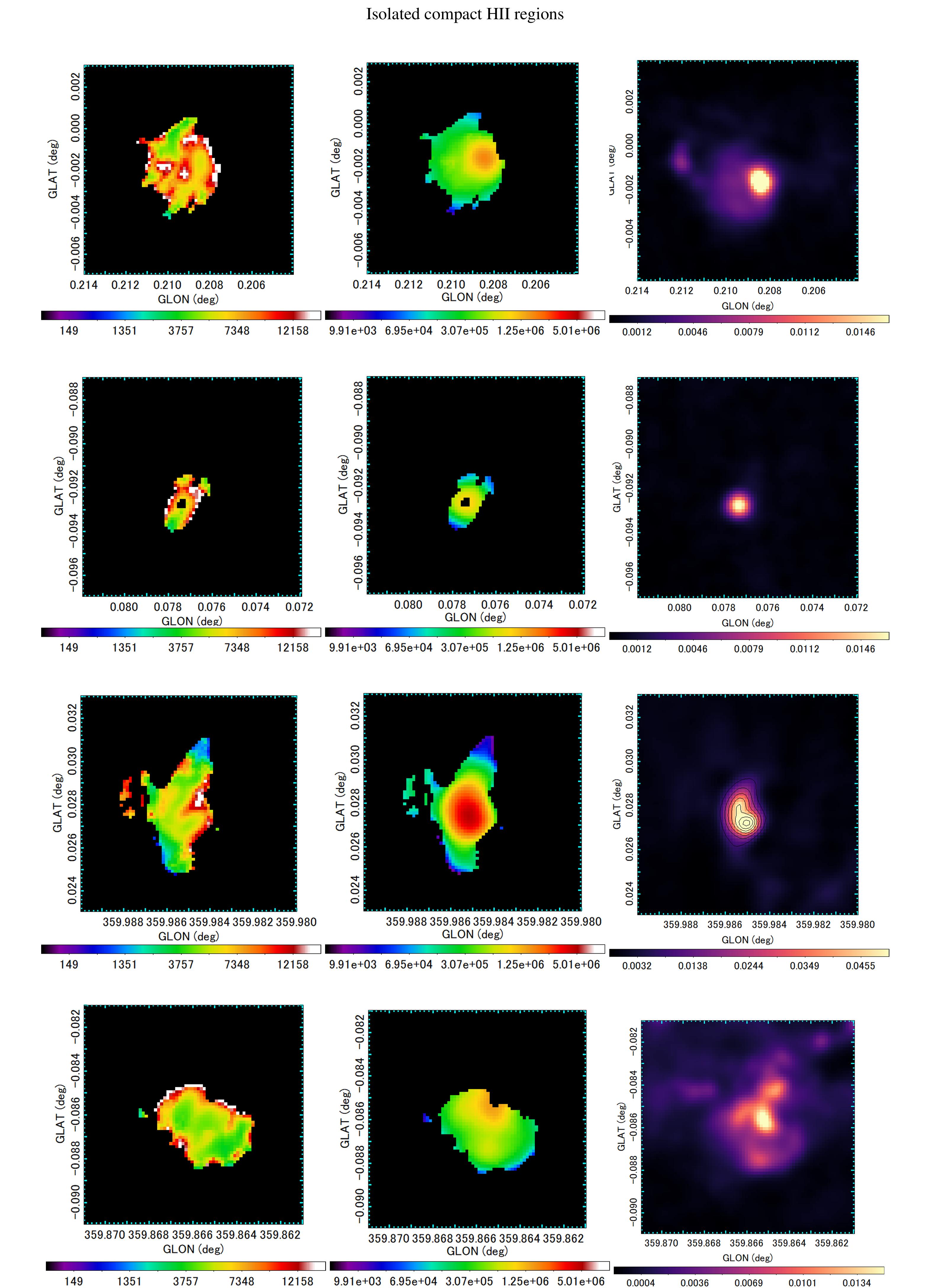} 
\end{center} 
\caption{Same as figure \ref{fig_indiv1}, but for isolated sources.
{Alt text: Same as figure \ref{fig_indiv1}.}
}
\label{fig_indiv3}  
\end{figure*}

In this section, we describe the individual regions based on the maps of $\Te$ and $EM$, which we show in figures \ref{fig_indiv1}, \ref{fig_indiv2}, and \ref{fig_indiv3}, where we enlarge the maps of electron temperature and emission measure for the individual regions labeled in figures \ref{fig-full} and \ref{fig-semiful}, along with the referenced continuum maps.
In table \ref{tab2} we list the measured values in representative regions on these maps.

\ss{Sgr B2}\label{ssSgrB2}

Sgr B2 is composed of many HII clumps centered on the central strong continuum source Sgr B2 Main \citep{M1993}.
Hydrogen recombination line analyses have been performed in the microwave ranges, resulting in electron temperature of $\sim 5000$ -- 10000 K \citep{M1993,P1996,Me2019,Me2022}.

We obtained consistently high value, $\Te=8926\pm 139$ K for the Main (table \ref{tab2}) with a high $EM \sim 7\times 10^7$ \emu.
Surrounding Sgr B2 HII regions have $\Te\sim 5000-5700$ K with $EM\sim 3\times 105 - 1\times 10^7$ \emu.

\begin{figure}   
\begin{center}        
\includegraphics[width=\lw]{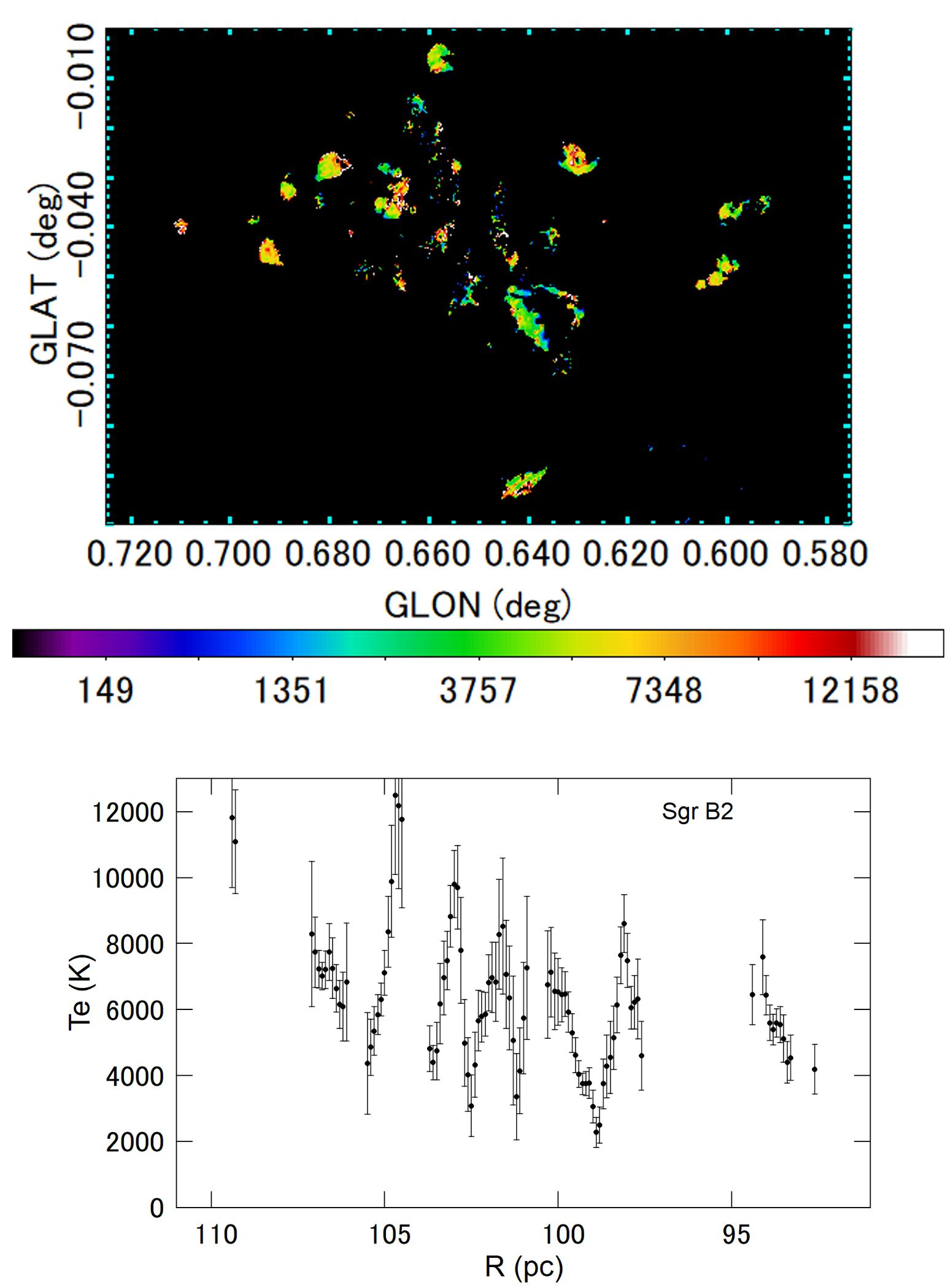} 
\end{center} 
\caption{
[Top] Electron temperature map of Sgr B2.
[Bottom] Averaged $\Te$ as a function of the "projected" distance $R$ from \sgrastar taken from table \ref{tab:TeCMZ}.
Note that the horizontal axis is reversed. Short and long bars represent standard error of the mean and standard deviation in the bin, respectively, and cross section across Main.
{Alt text: Electron temperature map of Sgr B2 and cross sections of $\Te$.Electron temperature map of Sgr B2.}
}
\label{fig_indiv_B2big}  
\end{figure}

In figure~\ref{fig_indiv_B2big} we show a close up of the whole Sgr B2 region with a plot of the averaged $\Te$ as a function of the projected distance $R$ from \sgrastar (approximately the longitude).

\begin{figure}   
\begin{center}           
\includegraphics[width=\lw]{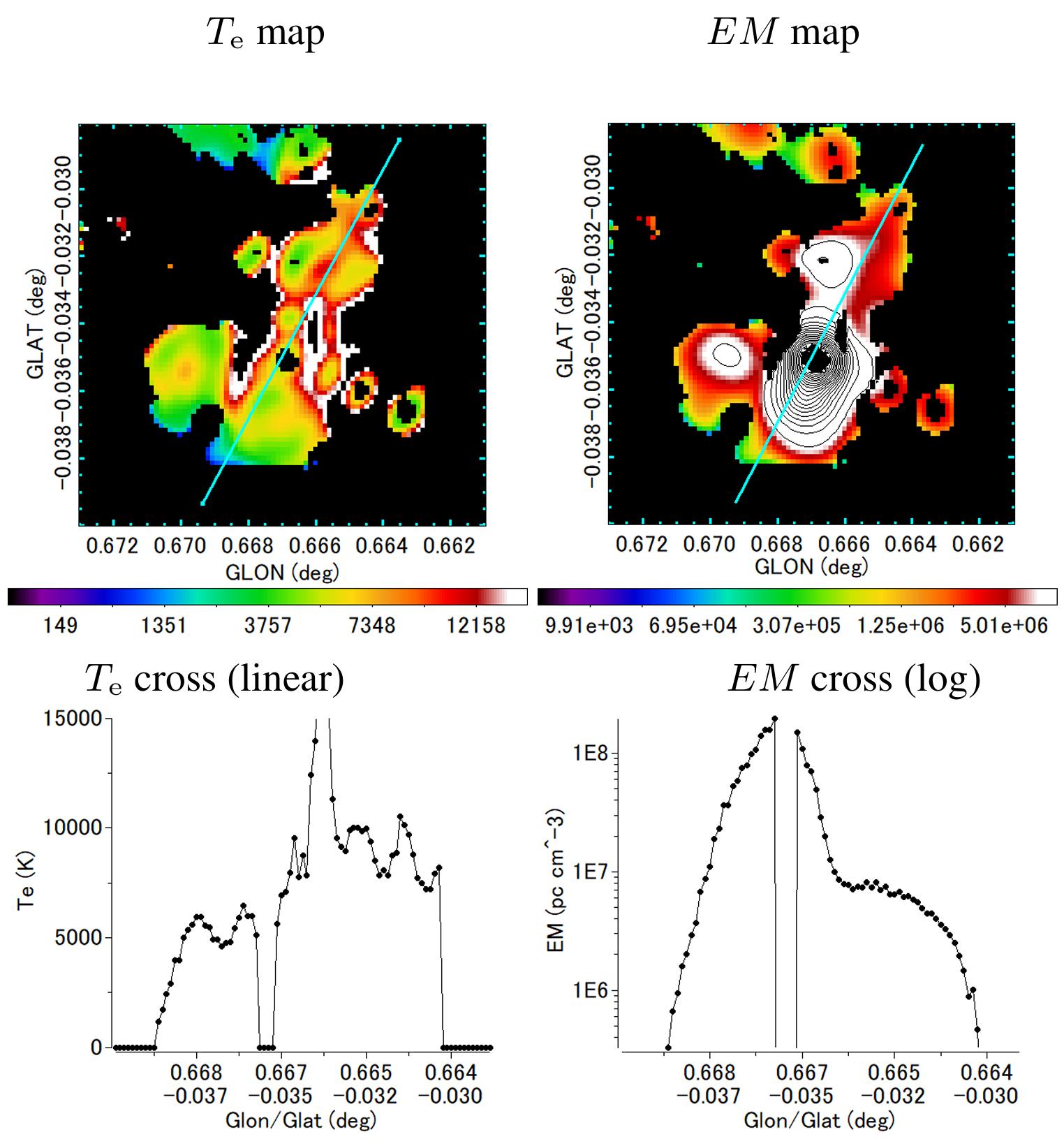} 
\end{center} 
\caption{
[Top] Electron temperature and emission measure maps of Sgr B2 Main. Contours are at every $10^7$ \emunit.
[Bottom] Cross sections of $\Te$ and $EM$ of Sgr B2 Main along the lines in the top panels.

{Alt text: Electron temperature and emission measure across Sgr B2 Main.}
}
\label{fig_indiv_B2_main}  
\end{figure}

{We further enlarge Sgr B2 Main in figure \ref{fig_indiv_B2_main} with a cross section along the lines shown in the top panels.
We find a steep $\Te$ gradient with $d\Te/dx\simeq 3000$ K pc$^{-1}$ and an extremely high peak value of $EM\sim 3 \times 10^8$ \emu.}
These figures demonstrate how strongly the electron temperature and emission measure vary from clump to clump, as well as within each clump.
For its diameter of $D\sim 5''.4 \sim 0.21$ pc, the peak $EM$ corresponds to an electron density as high as $\Ne\sim 4\times 10^4$ cm$^{-3}$.

It is interesting to note that $\Te$ is relatively stable at $\sim 6000-8000$ K near the sharp central peak of $EM$, whereas a high value of $EM\sim 2\times 10^8$ \emu is recorded towards the peak, which yields an electron density of $n_{\rm e}\sim 3\times 10^4$ cm$^{-3}$ for a half width of the $EM$ profile of $0\deg.004\sim 0.6 \epc$. 
This region has been resolved into $19$ ultracompact H II regions having $EM\sim 2\x 10^9$ \emu within a $\sim0.06$ pc diameter (not resolved here as our resolution is 0.1 pc) \citep{DeP1998}.   
\ss{Sgr B1}

Sgr B1 is an ensemble of medium temperature HII regions with an average value in the whole area of $\Te=4100 - 7800$ K, consistent with the earlier observations by \citet{M1992} who obtained $\Te\sim 5000$ K.
The emission measure is also uniform at $\sim 1-6\times 10^5$ \emunit.

\ss{Sickle and Pistol}  \label{secPistol}

The Sickle and Pistol have been extensively studied in the continuum and cm-wave recombination lines \citep{L1997} with the derived $\Te$ of $\sim5500$ and $\sim 3600$ K, respectively.

We obtain a warmer value, $\Te=5800-7900$ K, for Sickle, and the emission measure is uniform at $\sim 2\times 10^5$ \emunit.
The brightest region shows a lower $\Te \sim 5800$ K, consistent with the current measurements.

Pistol has a value of $\Te=4800$ K (table \ref{tab2}), close to the current observations in cm-wave lengths.
The emission measure is higher at $\sim 4\times 10^5$ \emunit.  
If we assume that the line-of-sight depth is comparable to the width, which is $L\sim 0.2$ pc for the Pistol, we obtain $\Ne\sim 2000$ cm$^{-3}$.

The ``smoke'' in the southwest of the Pistol is cool ($\sim 5000$ K), diffuse, and faint, with an emission measure as low as $\sim 10^5$ \emunit.
The continuum point source near the center of the pistol's arc is not detected in \h40.

The LVD (position-velocity diagram (PVD) in longitude) shown in figure \ref{fig-full} indicates that Sickle and Pistol have quite distinct radial velocities of $\vlsr \sim 40$ and $\sim120$ \kms, respectively, which are also reported in the current observations \citep{L1997}.
This velocity difference, as large as $\sim 80$ \kms, suggests that the two objects are kinematically distinct from each other; hence, they are significantly displaced along the line of sight, although both draw concentric arch-like structures in the sky.

If they are rotating with the CMZ disk of molecular gas at a rotation velocity of $\vrot \sim 110$ \kms \citep{S2025b}, the Pistol is located near the tangential point with a radius from \sgrastar of $R\sim 23$ pc, whereas the Sickle is at $R\sim 74$ pc.
These kinematically different locations suggest that at least one of the two objects, or possibly both, are not interacting with the Radio Arc.
  
\ss{Thermal filaments (Bridges)}

The thermal filaments (Bridges) appear as four parallel ridges, having $\Te\sim 4800-6000$ K.
These are consistent with the current measurements from VLA, showing 5000 to 6900 K in nearly the same regions \citep{L2001}.
Emission measure is relatively uniform at $EM\sim 2\times 10^6$ \emunit despite their large extent.
These regions have 'forbidden' radial velocities at $\vlsr \sim -20$ to $-60$ \kms, suggesting that they are moving on extremely eccentric orbits.
It is interesting to point out that these negative velocities are close to that of the 4-kpc expanding molecular ring in the Galactic disk at $R\sim 4 kpc$.

\ss{Sgr A HII regions}

From ALMA H42$\alpha$ line and continuum observations, an electron temperature of 
$\Te = 5150 - 5920$ K has been measured in the Sgr A HII regions \citep{T2019}.
Our measurement yields 
$\Te\sim 5300$ K in the north-east clump, 
$\sim 6900$ K in the companion to the north-east clump, 
$\sim 6500$ K in the south clump, and 
$\sim 4250$ K in the south-east clump.
The brightest north-east HII region has a high value of 
$EM\sim (1.46\pm0.04)\times 10^6 $ \emunit, while the other clumps have lower values of 
$EM\sim 6\times 10^5$ \emunit.

\ss{Minispiral and \star}

An extensive study has revealed detailed kinematics and physical properties of the Minispiral, showing that the western heavy arm has a high temperature of $\Te\sim 8000-14000$ K, whereas the eastern arm has a lower value of $\sim 6000-8000$ K \citep{T2019}.

\begin{figure*}
\begin{center}     
\includegraphics[width=.95\lw]{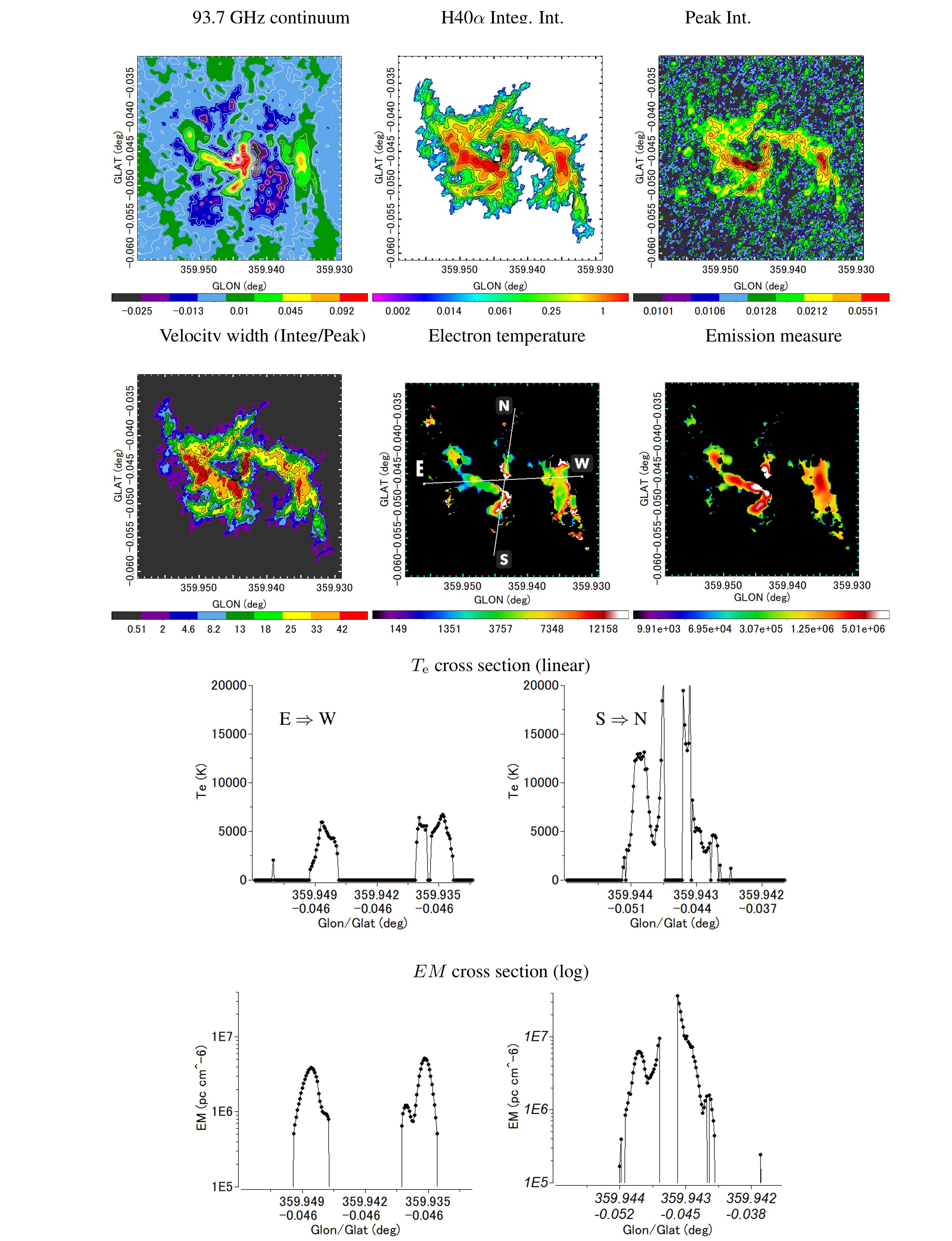}   \\
\end{center} 
\caption{Minispiral maps at [top left] 93.7 GHz continuum, 
[top middle] \h40 line integrated intensity, [top right] \h40 peak intensity {(see also figure \ref{fig:sidelobe}),}
[middle left] velocity width, [middle] electron temperature, $\Te$, and [middle right] emission measure $EM$.
The bottom panels show cross sections of $\Te$ and $EM$ across the E and W arms and along the S and N arms across \sgrastar as indicated by the white lines in the $\Te$ map. 
{Alt text:  Minispiral maps in continuum, \h40 integrated and peak intensity, velocity width, $\Te$, $EM$, and $\Te$ cross sections.}
}
\label{fig_mini} 
\end{figure*}

Figure \ref{fig_mini} shows the maps of the Minispiral from the present analysis and cross sections of $\Te$ and $EM$ in the EW and SN directions along the two white lines shown in the map (See also table \ref{tab2}).
The main ridge of the eastern arm is relatively cool at $\Te\sim$ 3400 -- 5200 K, while the western arm is moderate at 4500 -- 5300 K.
Both regions have high emission measures of $EM \sim 2\times 10^6$ \emunit.

On the other hand, the central vertical arm running from south to north (SN arm), closest to \sgrastar, has an extraordinarily high temperature of $\Te=15036 \pm 1321$ K.
The emission measure of the SN arm is as high as $EM\sim 2\times 10^6$ \emu, increasing toward the nucleus and reaching $\sim 3\times10^7$ \emu.

{It is interesting to point out that \star appears as a sharp absorption hole of \h40 peak intensity map as shown in figure \ref{fig:sidelobe}.
This fact indicates that the NS (closest) arm of Minispiral crosses the line of sight.
Its optical depth is 
\be
\tau\sim \delta I_{\rm H40\alpha}/I_{\rm Sgr~A^*}\sim 4\times 10^{-3}.
\ee
According to equation (\ref{eq:tau}) assuming $\Te\sim 12000$ K, the emission measure of this \H40 absorber is as large as $EM\sim 3\times 10^8$ \emu, two orders of magnitude greater than the value from TeEM analysis, $\sim 3\times 10^6$ \emu in the nearest NS arm.
The higher value by \star absorption may be due to the pin point measurement toward \star which captures a higher density clump of HII gas in the NS arm.}

These anomalously high temperatures and density in proximity to \sgrastar and the absence of spiral nature suggest that the excitation (ionization) mechanism of the NS arms may differ from that of the E and W spiral arms.
This may possibly be due to tidal compression and strong shear on the NS arms from the stronger gravity of the central massive black hole or some cause related to the amplified strong magnetic field \citep{M2026}.

\ss{Isolated HII regions}

We have identified four isolated HII regions, as shown in figure \ref{fig_indiv3}.
They are extended objects associated with compact continuum sources, often exhibiting shell-like structures.
The electron temperature ranges from $\sim 4600$ to 9400 K, and the emission measure ranges from $\sim 3\times 10^5$ to $2\times 10^6$ \emu.

\ss{Sgr C}

Sgr C was not included in the analysis because the \h40 integrated intensity map did not show significant emission to calculate the velocity width by dividing by the peak intensity.
For reference, we show the 2D maps of the 99.6 GHz continuum emission, \h40 peak intensity, and integrated intensity of the line around Sgr C in figure \ref{fig-full}.
This source needs to be analyzed individually from the data cube in further detail, but this is beyond the scope of this paper.

\ss{Mean $\Te$ and $EM$ in individual regions}
We measured the mean $\Te$ as well as $EM$ in individual HII regions presented in figures \ref{fig_indiv1} – \ref{fig_indiv3}, along with the standard deviation (SD), and listed the results in table \ref{tab2}
Thereby, we took the mean within a circle of radius $r$, with the diameter fitting the apparent minor axis size. 
Radius $r$ was further used to calculate the electron density $\Ne$ from $EM$, where the line of sight size of the object is assumed to be the apparent width, so that $L\sim 2r$.
\be
\Ne\sim \sqrt{EM/(2r)}.
\ee

\begin{table*}[]
    \caption{Individual means in HII regions of electron temperature $\Te$, emission measure $EM$, and electron density $\Ne$.}
    \centering
     \label{tab2}
    \begin{tabular}{cccccccc}
    \hline
    \hline
Source  &$(l,b)$ &$\Te~(\pm {\rm SD})$& $EM ~(\pm {\rm SD})$& $\Ne$& $L\sim$Diameter  \\ 
&(deg)& (K)& ($10^6$ \emunit)&(cm$^{-3}$) &(pc)\\
\hline

Sgr B2 & G   0.692  -0.045 & $    6582.3\pm      90.3$ & $     3.231\pm     0.066$ & $    4485.0\pm      48.3$ &     0.32\\
 & G   0.680  -0.028 & $    6419.7\pm     270.2$ & $    16.313\pm     0.004$ & $    8718.3\pm       7.4$ &     0.41\\
 & G   0.667  -0.036 & $    8926.3\pm     138.8$ & $    68.634\pm     0.244$ & $   19210.7\pm      48.0$ &     0.36\\
 & G   0.658  -0.007 & $    5948.0\pm      12.3$ & $     0.527\pm     0.009$ & $    1657.6\pm      12.9$ &     0.36\\
 & G   0.642  -0.092 & $    6723.2\pm      31.5$ & $     0.300\pm     0.001$ & $    1280.2\pm       2.7$ &     0.36\\
 & G   0.640  -0.059 & $    3814.7\pm       3.4$ & $     0.415\pm     0.006$ & $    1548.6\pm      10.7$ &     0.34\\
 & G   0.631  -0.026 & $    8789.4\pm     110.9$ & $     1.339\pm     0.009$ & $    2781.2\pm       8.7$ &     0.34\\
 & G   0.600  -0.048 & $    4921.4\pm      83.4$ & $     2.063\pm     0.030$ & $    3958.1\pm      29.8$ &     0.26\\
 & G   0.562  -0.044 & $    4533.4\pm     111.1$ & $     0.236\pm     0.000$ & $    1339.7\pm       1.4$ &     0.26\\
\hline
Sgr B1 & G   0.528  -0.085 & $    5753.7\pm     146.1$ & $     0.336\pm     0.009$ & $    1438.7\pm      21.6$ &     0.32\\
 & G   0.515  -0.048 & $    4129.9\pm      43.3$ & $     0.130\pm     0.002$ & $     869.6\pm       5.6$ &     0.34\\
 & G   0.489  -0.073 & $    7862.9\pm     194.2$ & $     0.591\pm     0.006$ & $    1610.7\pm       8.7$ &     0.45\\
 & G   0.489  -0.057 & $    4843.4\pm      24.4$ & $     0.276\pm     0.000$ & $    1094.0\pm       1.2$ &     0.45\\
 & G   0.471  -0.056 & $    5943.0\pm       9.2$ & $     0.347\pm     0.024$ & $    1525.9\pm      47.7$ &     0.29\\
\hline Sickle & G   0.197  -0.064 & $    7882.9\pm      91.6$ & $     0.198\pm     0.003$ & $    1172.2\pm       8.4$ &     0.29\\
 & G   0.188  -0.052 & $    5782.1\pm     216.3$ & $     0.168\pm     0.001$ & $    1081.4\pm       3.1$ &     0.29\\
\hline
Pistol & G   0.162  -0.063 & $    4746.4\pm      76.5$ & $     0.423\pm     0.006$ & $    1991.3\pm      14.7$ &     0.21\\
 & G   0.160  -0.068 & $    4834.1\pm     189.1$ & $     0.097\pm     0.002$ & $    1010.4\pm       8.7$ &     0.19\\
\hline Bridge & G   0.097   0.020 & $    4936.6\pm      63.6$ & $     0.158\pm     0.005$ & $    1016.3\pm      16.3$ &     0.30\\
 & G   0.076   0.012 & $    6719.4\pm      49.7$ & $     0.193\pm     0.001$ & $    1132.3\pm       3.9$ &     0.30\\
 & G   0.069   0.041 & $    4771.3\pm      65.2$ & $     0.244\pm     0.002$ & $    1266.8\pm       5.9$ &     0.30\\
 & G   0.042   0.020 & $    6884.7\pm     204.7$ & $     0.327\pm     0.007$ & $    1284.3\pm      14.0$ &     0.38\\
\hline Sgr A HII & G  -0.019  -0.074 & $    5263.9\pm      67.1$ & $     1.464\pm     0.041$ & $    2945.0\pm      46.4$ &     0.33\\
 & G  -0.022  -0.077 & $    6870.4\pm     269.2$ & $     0.570\pm     0.008$ & $    2076.9\pm      14.7$ &     0.26\\
 & G  -0.026  -0.081 & $    6493.2\pm      67.0$ & $     0.534\pm     0.015$ & $    2014.3\pm      30.4$ &     0.26\\
 & G  -0.033  -0.081 & $    4251.6\pm     432.0$ & $     0.829\pm     0.007$ & $    3049.4\pm      13.3$ &     0.18\\
\hline Minispiral & G  -0.048  -0.042 & $    5208.3\pm      36.5$ & $     1.312\pm     0.051$ & $    4252.4\pm      87.4$ &     0.14\\
 & G  -0.052  -0.046 & $    3418.9\pm      16.9$ & $     2.011\pm     0.033$ & $    5503.8\pm      46.1$ &     0.13\\
 & G  -0.064  -0.046 & $    5276.3\pm     179.6$ & $     2.331\pm     0.014$ & $    4379.9\pm      11.7$ &     0.24\\
 & G  -0.065  -0.049 & $    4526.6\pm       1.2$ & $     1.677\pm     0.059$ & $    3866.3\pm      64.5$ &     0.22\\
 & G  -0.056  -0.050 & $    8478.9\pm     128.6$ & $     3.113\pm     0.000$ & $    8516.9\pm       0.8$ &     0.09\\
 & G  -0.057  -0.045 & $   15036.2\pm    1321.0$ & $    16.767\pm     1.633$ & $   23139.7\pm    1181.3$ &     0.06\\
 & G  -0.057  -0.044 & $    4760.0\pm     119.6$ & $     3.921\pm     0.278$ & $    8647.8\pm     326.4$ &     0.10\\
\hline Isolated & G   0.209  -0.002 & $    7686.6\pm     111.6$ & $     0.604\pm     0.006$ & $    2040.6\pm      11.1$ &     0.29\\
 & G   0.077  -0.092 & $    9417.4\pm     708.7$ & $     0.344\pm     0.011$ & $    2419.6\pm      41.0$ &     0.12\\
 & G  -0.015   0.028 & $    6500.1\pm     710.2$ & $     1.849\pm     0.044$ & $    3843.0\pm      47.5$ &     0.25\\
 & G  -0.135  -0.087 & $    4617.3\pm      42.9$ & $     0.568\pm     0.015$ & $    2147.9\pm      28.2$ &     0.24\\
         \hline
    \end{tabular}
\end{table*}

\section{Discussion}  

\ss{General remarks}

We presented a method called TeEM ($\Te, ~EM$ Mapping) to derive the electron temperature and emission measure using three 2D maps; continuum emission, recombination peak line intensity, and integrated line intensity maps.
Applying the method to the ACES survey of the 99.6 GHz continuum and \h40 line data, we obtained distribution maps of $\Te$ and $EM$ in the CMZ, as presented in figures \ref{fig-full} to \ref{fig_indiv3}. 

Based on the maps, we showed that the electron temperature is relatively uniform over the CMZ, around \Tcmztext. 

The value is consistent within the uncertainties with the current measurements of the individual HII regions and thermal filaments using the VLA and ALMA inside the CMZ \citep{M1986,M1992,M1993,Z1993,P1996,L1997,L2001,T2017,T2019}.
{It is also consistent with a slightly lower value at $\sim 4360\pm 900$ K observed in the GCL above the CMZ \citep{Nagoshi+2019}.}

\ss{Galactic $\Te$ gradient and metallicity}

In figure \ref{fig:te-r} we plot the derived electron temperature for the HII regions in CMZ as a function of the distance from \sgrastar as listed in table \ref{tab:TeCMZ}.
Although individual HII regions show large scatter, the averaged $\Te$ is nearly constant at $\sim 6000$ K.

\begin{figure}   
\begin{center}  
\includegraphics[width=\lw]{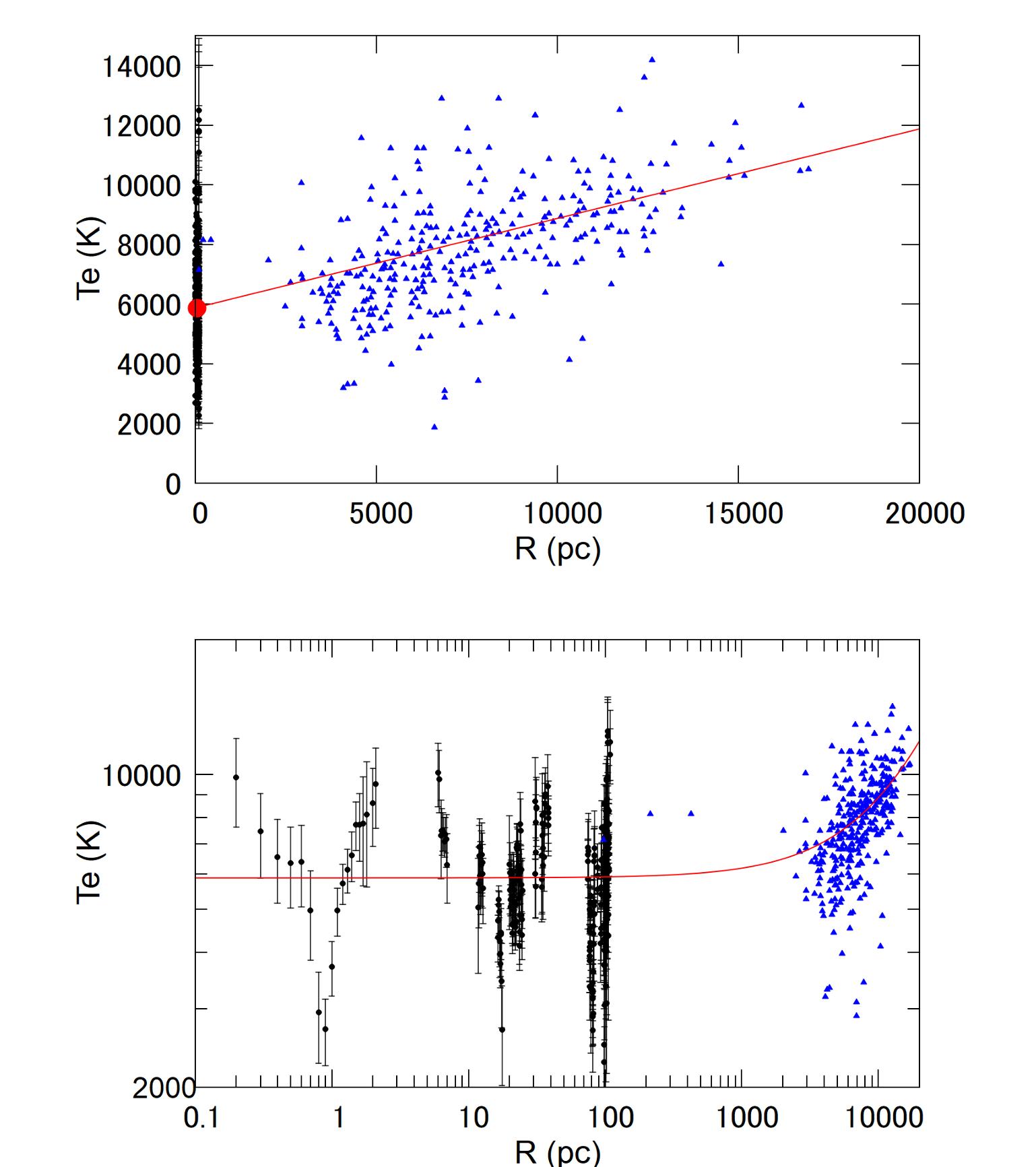}   
\end{center} 
\caption{Same as figure \ref{fig:te-r}, but with $\Te$ in the whole Milky Way taken from \citet{K2024} by triangles in linear (top) and logarithmic (bottom) scalings. The lines represents a linear relation between $\Te$ and $R$ given by equation \ref{eq:TeR}.
{Alt text: Electron temperature as a function of the Galacto-centric distance.}}
\label{te-r-mw}  
\end{figure}

Combining the $\Te$ plot in the CMZ with a wider Galactic disk study \citep{S1983,B2011,K2024}, we created a $\Te$ plot as a function of $R$ from GC to the outer Milky Way \citep{B2011,K2024} as shown in figure \ref{te-r-mw}.  
The plots show that $\Te$ is nearly flat from the center to $R\sim 3$ kpc, as shown by the lower panel (logarithmic plot).
It then increases from the GC value  up to $\sim 10^4$ K in the outer regions of the Galactic disk at a rate of $d\Te/dR\simeq +300$ K kpc$^{-1}$, as indicated by the red line in figure \ref{te-r-mw}.
This can be approximated by a linear relation as
\be 
\Te \sim 5872+ 300(R/ {\rm kpc}) ~[K]. \label{eq:TeR}
\ee 
 The $\Te$--$R$ plots will provide new insights into the metallicity gradient issue in the Galactic disk \citep{S1983,MD2023,M2026} including the GC.
For example, we may rewrite the empirical metal-to-$\Te$ relation \citep{M2026} to a metal-to-$R$ relation as
\be
12+\log (\rm{O/H})=9.29_{\pm0.02}-0.96_{\pm0.02}( \Te/10^4{\rm K}) \nonumber
\ee  
\be 
\sim 8.72-0.029 (R/\ekpc).
\label{eqZ}
\ee
{We point out that these relationships and $\Te$-$R$ diagrams in figure \ref{te-r-mw} can be also used to constrain the GC distance of an HII region. For example, a low $\Te$ measured in GCL by \citet{Nagoshi+2019} may indicate that the gas has a high metallicity and is therefore located inside $R\lesssim 3$ kpc, consistent with their distance estimation, $R\sim 2.5$ kpc. }

\ss{Internal $\Te$ variation within an HII region}

In some HII regions, the electron temperature varies smoothly and steeply across the region. 
A typical example is observed in Sgr B2 Main, as shown in figure \ref{fig_indiv_B2_main}, where $\Te$ varies from $\sim 4000$ K at the eastern edge to $\sim 9000$ K at the western edge within 1.3 pc, with a gradient as steep as $d\Te/dx\sim 4400$ K pc$^{-1}$. 
If the metallicity-to-electron temperature relation \citep{M2026} is applied on such a small scale ($\sim 1$ pc), the gradient in Sgr B2 Main corresponds to $d(12+\log(O/H))/dx\sim -0.48$ dex pc$^{-1}$, requiring that the metallicity in Sgr B2 Main varies by 0.6 dex over 1.1 pc across the major axis.
However, such a small scale variation of the metallicity is not realistic. 
We, therefore, consider the steep $\Te$ gradient and variation of $\Te$ in individual HII regions to be due to physical and structural characteristics within the regions, such as knotty ionizing fronts created by surrounding molecular clumps \citep{O2003}. 

\begin{figure}   
\begin{center}  
\includegraphics[width=\lw]{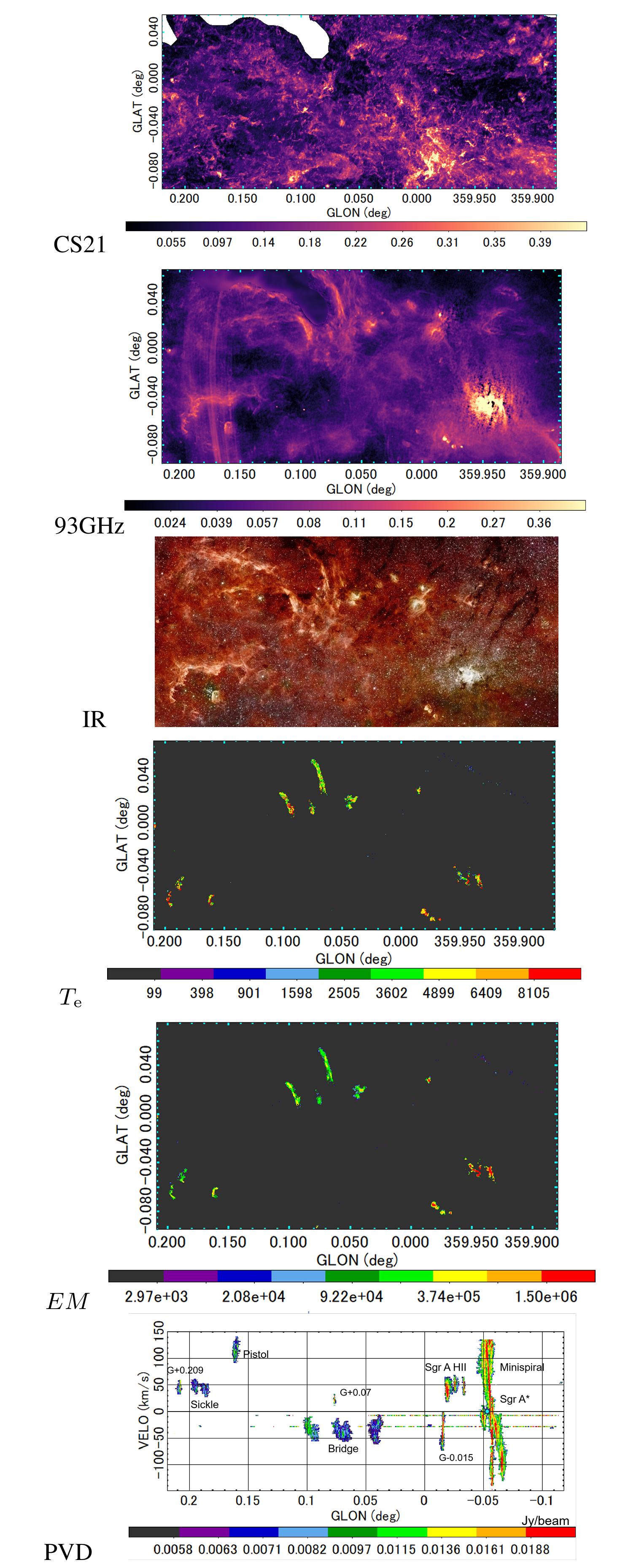}  
\end{center} 
\caption{
{Comparison of the results on $\Te$ and $EM$ with the infrared view of the central region : 
Red -- NICMOS 187 \mum (sensitive primarily to Paschen-$\alpha$ \citep{wang+2010}) + IRAC 8.0 \mum (hot dust); Orange -- IRAC 5.8 \mum (stars and very hot dust); Green -- IRAC 4.5 microns (stars); Blue -- IRAC 3.6 \mum (stars).
\tiny
https://science.nasa.gov/missions/hubble/an-infrared-view-of-the-galaxy/. 
Image Credits: Hubble: NASA, ESA, and Q.D. Wang (University of Massachusetts, Amherst); Spitzer: NASA, Jet Propulsion Laboratory, and S. Stolovy (Spitzer Science Center/Caltech). 
{Alt text: Comparison with infrared view of the central region.} 
}}
\label{fig:nasa}  
\end{figure}

{
\ss{Comparison with other observations}
\sss{Infrared maps}
Figure \ref{fig:nasa} compares our ACES $\Te$ and $EM$ maps of the HII regions in the central region with a \cs~ peak intensity map, showing molecular gas, a 93.7 GHz continuum map combined with single-dish image, showing free-free emission, with a color-composite infrared image \citep{wang+2010,Dong+2011}, showing 
NICMOS 187 \mum (Paschen-$\alpha$) + IRAC 8.0 \mum (hot dust); IRAC 5.8 \mum (stars and very hot dust); IRAC 4.5 \mum (stars); IRAC 3.6 \mum (stars). 
The ACES data primarily capture the brightest, sharply defined infrared structures, as highlighted by the Paschen-$\alpha$ map, associated with bright clusters of red giants. 
The highly localized $\Te$ and $EM$ distribution is partly due to the fact that radio interferometers are insensitive to diffuse emission. 
The key benefit of using radio RRL observations to study ionized gas is their immunity to dust extinction along with the determination of $\Te$ and $EM$. 
A systematic comparison between the infrared and millimeter data could be used to infer the extinction toward these structures, but such an analysis is beyond the scope of this paper.
\sss{Molecular lines}
A glance at the \cs ~line map in the upper panel reveals that none, except for Sgr A HII, of the star clusters exciting the HII regions are associated with their parent molecular cloud.
This suggests that Schmidt's law no longer applies in the scale discussed in these maps within $R\lesssim 30$ pc. 
This further poses a problem: unlike typical SF regions in galactic disks, it is difficult to kinematically analyze the motion of SF regions, hence to determine the location.
\sss{PVD (LVD) Kinematics} 
The problem of kinematics of SF regions may be solved by measuring the velocity of HII gas using RRL.
Although a thorough consideration will be given in a separate paper, we here comment on a possible use of RRL PVDs (figures \ref{fig-full}, \ref{fig:nasa} ) that reveal:
\begin{itemize}
\item Fast rotation of the Minispiral in the regular Galactic rotation, as already studied in detail  \citep{S2026cnd}.
\item The Sickle 
($\Vlsr\sim 45$ \kms) and Pistol 
($\sim 120$ \kms) are distinct objects with their kinematical distances from \sgrastar of $R\sim 24$ and 75 pc, respectively. At least one of these objects may not be related to the non-thermal filaments of the Radio Arc.
\item Thermal filaments (Bridges) have forbidden radial velocities, counter-rotating around \sgrastar, posing a doubt whether they are inside the CMZ.
\item  G-0.014+0.028 has a forbidden velocity at $\vlsr \sim -35$ \kms and is counter rotating at $d\vlsr / dl \sim -3\times 10^4$ \kms deg$^{-1}$ $=200$ \kms pc$^{-1}$.
Such a rapid counter rotation is difficult to explain by contraction of a cloud in the nuclear disk in Galactic rotation.
\end{itemize} 
}

\section{Summary}
\label{summary}

We obtained maps of the electron temperature $\Te$ and emission measure $EM$ of HII regions in the CMZ by applying the TeEM method to ACES maps of the 99.6 GHz continuum intensity, \h40 line peak intensity, and integrated intensity. 
We presented the maps of the whole CMZ as well as detailed $\Te$ and $EM$ maps for individual HII regions and discussed their properties, highlighting Sgr B2, the Pistol, and the minispiral.
We found that the global electron temperature is relatively constant at \Tcmztext inside the CMZ, while individual HII regions exhibit a higher order variation across each region. The highest value of $\Te\sim 12000-18000$ K was recorded toward the Minispiral along the North-South arm closest to \sgrastar.
The $EM$ distribution is more diverse, ranging from $\sim 10^5$ to $\sim 10^8$ \emunit, varying from region to region as well as within individual regions. 
The highest value of $EM\sim 2\times 10^8$ \emu was recorded towards Sgr B2 Main.
 
{\tiny
\scriptsize{
\begin{ack}
This paper makes use of the following ALMA data: ADS/JAO.ALMA\#2021.1.00172.L. 
ALMA is a partnership of ESO (representing its member states), NSF (USA) and NINS (Japan), together with NRC (Canada), MOST and ASIAA (Taiwan), and KASI (Republic of Korea), in cooperation with the Republic of Chile. The Joint ALMA Observatory is operated by ESO, AUI/NRAO and NAOJ.
 The data analysis in this paper was performed at the Astronomical Data Center of the National Astronomical Observatories of Japan.\\
Y.S. acknowledges support from JSPS KAKENHI Grant Number JP24H00004.\\
%
C.B. acknowledges  funding  from  NSF under Award  Nos. 2108938, 2206510, and CAREER 2145689, as well as from the National Aeronautics and Space Administration through the Astrophysics Data Analysis Program under Award ``3-D MC: Mapping Circumnuclear Molecular Clouds from X-ray to Radio,” Grant No. 80NSSC22K1125.\\
A.G. acknowledges support from the NSF under AAG 2206511 and CAREER 2142300.\\
X.L. acknowledges support from the Strategic Priority Research Program of the Chinese Academy of Sciences (CAS) Grant No.\ XDB0800300, the National SKA Program of China (2025SKA0140100), the National Natural Science Foundation of China (NSFC) through grant Nos.\ 12273090 and 12322305, and the National Key R\&D Program of China (No.\ 2022YFA1603101).\\
F.N.L gratefully acknowledges financial support from grant PID2024-162148NA-I00, funded by MCIN/AEI/10.13039/501100011033 and the European Regional Development Fund (ERDF) “A way of making Europe”, from the Ramón y Cajal program (RYC2023-044924-I) funded by MCIN/AEI/10.13039/501100011033 and FSE+, and from the Severo Ochoa grant CEX2021-001131-S, funded by MCIN/AEI/10.13039/501100011033.\\
%
%
L.C., V.M.R. and I.J.-S. acknowledge support from the grant PID2022-136814NB-I00 by the Spanish Ministry of Science, Innovation and Universities/State Agency of Research MICIU/AEI/10.13039/501100011033 and by ERDF, UE. 
The project that gave rise to these results received the support of a fellowship from the "la Caixa" Foundation (ID 100010434). The fellowship code is LCF/BQ/PR25/12110012. 
V.M.R. also acknowledges support from the grant RYC2020-029387-I funded by MICIU/AEI/10.13039/501100011033 and by "ESF, Investing in your future", from the Consejo Superior de Investigaciones Cient{\'i}ficas (CSIC) and the Centro de Astrobiolog{\'i}a (CAB) through the project 20225AT015 (Proyectos intramurales especiales del CSIC); and from the grant 
CNS2023-144464 funded by MICIU/AEI/10.13039/501100011033 and by “European Union NextGenerationEU/PRTR”.\\
I.J.-S. acknowledges support from ERC grant OPENS, GA No. 101125858, funded by the European Union.\\
P. Garc\'ia is sponsored by the Chinese Academy of Sciences (CAS), through a grant to the CAS South America Center for Astronomy (CASSACA). \\
%
%
D. Riquelme-V\'asquez acknowledges the financial support of DIDULS/ULS, through the project PAAI 2023.\\
J.W. gratefully acknowledges funding from the National Science Foundation under Award Nos. 2108938 and 2206510.\\
A.S.-M.\ acknowledges support from PID2023-146675NB-I00 (MCI-AEI-FEDER, UE) grant funded by MCIN/AEI/10.13039/501100011033 and by the European Union `Next GenerationEU’/PRTR, as well as the program Unidad de Excelencia María de Maeztu CEX2020-001058-M, and by the MaX-CSIC Excellence Award MaX4-SOMMA-IC.\\
Q.D.W.\ acknowledges support from NASA via grant GO3-24120X.\\
Part of this research was carried out at the Jet Propulsion Laboratory, California Institute of Technology, under a contract with the National Aeronautics and Space Administration (80NM0018D0004). D.C.L. acknowledges financial support from the National Aeronautics and Space Administration (NASA) Astrophysics Data Analysis Program (ADAP).\\
\end{ack}
  

\section*{Data availability} 
The interferometer data were taken from the survey data from ALMA cycle 8 Large Program "ALMA Central Molecular Zone Exploration Survey" (ACES, 2021.1.00172.L).
 
\section*{Conflict of interest}
The authors declare that there is no conflict of interest.
  }

\end{document}